\documentclass[12pt,notitlepage,aps,preprintnumbers,superscriptaddress,nofootinbib,aps,prd,floatfix]{revtex4-1}
\pdfoutput=1
\usepackage{graphicx}% Include figure files
\usepackage{dcolumn}% Align table columns on decimal point
\usepackage{bm}% bold math
\usepackage{slashed}
\usepackage{physics}
\usepackage{extarrows}
\usepackage{multirow,array}
\usepackage{amsmath,amssymb,physics}
\usepackage{subfigure,rotating}
\usepackage{enumitem}
\setlist{nolistsep}
\usepackage{xcolor}
\usepackage{url}
\usepackage{slashed}
\usepackage{hyperref}
\definecolor{nicered}{rgb}{0.5,0.,0.}
\definecolor{nicegreen}{rgb}{0.,0.5,0.}
\definecolor{niceblue}{rgb}{0.,0.,0.5}
\hypersetup{colorlinks,citecolor=nicegreen,linkcolor=nicered,urlcolor=niceblue}
\usepackage{orcidlink}
\usepackage{braket}
\usepackage{colortbl}
\usepackage{gensymb}%degree
\usepackage{comment}
\renewenvironment{eqnarray}{\begin{equation}\begin{aligned}}
{\end{aligned}\end{equation}}

\newcommand{\GeV}{\textrm{GeV}}

\newcommand{\cA}{\vartriangleright}
\newcommand{\cB}{\vartriangleleft}

\begin{document}
\preprint{INT-PUB-24-044, MSUHEP-24-015, SMU-PHY-24-05}
\title{General mass variable flavor number scheme for $Z$ boson production in association with a heavy quark at hadron colliders}
\author{Marco Guzzi\,\orcidlink{0000-0003-3430-2691}}
\email{mguzzi@kennesaw.edu}
\affiliation{Department of Physics, Kennesaw State University, Kennesaw, GA 30144, USA}
\author{Pavel Nadolsky\,\orcidlink{0000-0003-3732-0860}}
\email{nadolsky@smu.edu}
\affiliation{Department of Physics and Astronomy,
Michigan State University, 
East Lansing, MI 48824, USA\looseness=-1}
\affiliation{Department of Physics, Southern Methodist University, Dallas, TX 75275-0181, USA}
\author{Laura Reina\,\orcidlink{0000-0002-9268-5187}}
\email{reina@hep.fsu.edu}
\affiliation{Physics Department, Florida State University, Tallahassee, FL 32306-4350, USA}
\author{Doreen Wackeroth\,\orcidlink{0000-0001-8683-8338}}
\email{dw24@buffalo.edu}
\affiliation{Department of Physics, University at Buffalo, The State University of New York, Buffalo,
New York 14260-1500, USA}
\author{Keping Xie\,\orcidlink{0000-0003-4261-3393}}
\email{xiekeping@pitt.edu}
\affiliation{Department of Physics and Astronomy,
Michigan State University, 
East Lansing, MI 48824, USA\looseness=-1}

\date{\today}
\begin{abstract}
We present a methodology to streamline implementation of massive-quark radiative contributions in calculations with a variable number of active partons in proton-proton collisions. The methodology introduces \textit{subtraction} and \textit{residual} heavy-quark parton distribution functions (PDFs) to implement calculations in the Aivazis--Collins--Olness--Tung (ACOT) factorization scheme and its simplified realization in various processes up to the next-to-the-next-to-leading order in the QCD coupling strength. Interpolation tables for bottom-quark subtraction and residual distributions for CT18 NLO and NNLO PDF ensembles are provided in the common LHAPDF6 format. A numerical calculation of $Z$-boson production with at least one $b$ jet at the Large Hadron Collider beyond the lowest order in QCD is considered for illustration purposes.
\end{abstract}

\maketitle
\tableofcontents

\section{Introduction}
%%%%%%%%%%%%%%%%%%
The large inflow of high-precision measurements from the LHC and modern global QCD analyses to determine parton distribution functions (PDFs) of the proton require precise and accurate theory predictions for standard candle processes at higher orders in perturbation theory. Many factors compete in magnitude with higher-order radiative corrections in perturbative QCD calculations. Among those factors, heavy-quark (HQ) contributions can be significant, and HQ dynamics may affect the cross section calculation for a large variety of important processes at the Large Hadron Collider (LHC) and future colliders.

With the recent progress in theory calculations beyond next-to-next-to-leading order (NNLO) in the QCD strong coupling (e.g., N$^3$LO DIS ~\cite{Vermaseren:2005qc,Moch:2004xu,Moch:2007rq,Moch:2008fj,Kawamura:2012cr,Davies:2016ruz,Blumlein:2022gpp}, DIS with massive quarks~\cite{Ablinger:2024qxg},
N$^3$LO Drell--Yan~\cite{Baglio:2022wzu,Duhr:2020sdp,Duhr:2021vwj,Chen:2021vtu,Chen:2022lwc}, N$^3$LO Higgs production~\cite{Anastasiou:2015vya,Mistlberger:2018etf,Dulat:2018bfe,Dreyer:2016oyx,Cieri:2018oms}, and jet production in DIS~\cite{Gehrmann:2018odt,Currie:2018fgr}), consistent inclusion of HQ contributions in the cross section calculation for particle reactions is crucial for advancing on the precision frontier in the Run-3 of the LHC and beyond.  

As it is well known, various approaches or \textit{schemes} have been developed to simplify computations of HQ radiative contributions depending on the relative size of the HQ mass ($m_Q$) with respect to the hard scale of the process ($Q$). Two classical approaches, widely adopted since the early days of QCD, differ in their choices for keeping the HQ mass in the short-distance radiative contributions and for including the heavy quarks as active flavors in the renormalized QCD coupling strength $\alpha_s$ and PDFs.  

The classical {\it massive} scheme applies when the HQ mass $m_Q$ is approximately of the same size as the typical hard scale $Q$ of the process under consideration, and both these quantities are much larger than the mass of the proton $m_P$:
$Q^2\approx m_Q^2 \gg m_P^2$. This means that HQs are created solely as short-lived final states in high-$Q$ interactions, and the hard cross section retaining full mass dependence gives the correct description at energies comparable to the HQ mass threshold. In this framework, there is no HQ PDF. Instead, all HQ contributions belong to short-distance scattering, and $m_Q$ serves as a natural infrared cutoff. Renormalization in the classical massive scheme does not treat the HQs as active flavors, so such schemes are typically referred to as \textit{fixed flavor number} (FFN) schemes.

The classical {\it massless} scheme is valid when the typical scale of the process is much larger than both the HQ mass and the proton mass: $Q^2 \gg m_Q^2 \gg m_P^2$. In this case, the heavy quark is considered as essentially massless and also enters the running of the strong coupling $\alpha_s$. In addition, large logarithmic terms of the type $\ln^n(\mu^2/m_Q^2)$ in the corresponding partonic cross sections affect the perturbative convergence of the fixed-order calculation. They can be conveniently factorized by introducing an HQ PDF and resummed via Dokshitzer--Gribov--Lipatov--Altarelli--Parisi (DGLAP) equations. In the classical massless scheme, the number of active flavors $N_f$ increases with the typical energy scale, hence this approach is known as the \textit{zero-mass} (ZM) \textit{variable flavor number} (VFN) scheme.

Many high-precision observables currently included in global QCD analyses at NNLO (and more recently at partial N$^3$LO \cite{McGowan:2022nag,NNPDF:2024nan}) extend over a wide kinematic region of momentum fraction $x$ and momentum transfer $Q$.
It is therefore natural to evaluate all fitted cross sections in a factorization scheme that incorporates features of both
classical HQ schemes. Indeed, calculations performed in these schemes, when considered to all orders of perturbative QCD, correspond to just different ways of organizing the perturbative QCD series~\cite{Collins:1986mp,Barnett:1987jw,Olness:1987ep}. A more complete framework also interpolates between these two calculations in the appropriate kinematic regimes,
while assuring continuity of theoretical predictions and avoiding double counting of scattering contributions. Such interpolation can be performed in different ways, which are generically referred to as \textit{general-mass variable flavor number} (GMVFN) schemes. These schemes must converge reliably to the FFN scheme near the HQ production threshold. In GMVFN schemes, the number of quark flavors changes with the energy according to the quantum field theory treatments described in the seminal works of Appelquist and Carrazzone~\cite{Appelquist:1974tg}, and of Collins, Wilczek, and Zee (CWZ)~\cite{Collins:1978wz}.   

Several GMVFN factorization schemes have been proposed over time~\cite{Aivazis:1993kh,Aivazis:1993pi,Buza:1996wv,Thorne:1997ga,Collins:1998rz,Cacciari:1998it,Tung:2001mv,Kramer:2000hn,Kniehl:2004fy,Kniehl:2005mk,Alekhin:2009ni,Forte:2010ta,Guzzi:2011ew,Stavreva:2012bs,Kusina:2013slm} to interpolate between the massive FFN and zero-mass VFN schemes. 
In this work, we discuss an application of the GMVFN scheme known as Aivazis--Collins--Olness--Tung (ACOT)~\cite{Aivazis:1993kh,Aivazis:1993pi} to the case of the production of a $Z$ boson with at least one heavy quark in proton-proton collisions, and compare this approach to its simplified version as implemented in the Simplified-ACOT (S-ACOT) scheme for the same process~\cite{Tung:2001mv,Kramer:2000hn,Guzzi:2011ew}.
Variants of the ACOT/S-ACOT schemes have been successfully applied to account for heavy-flavor dynamics in DIS (S-ACOT-$\chi$ at NNLO for neutral~\cite{Guzzi:2011ew} and charged currents~\cite{Gao:2021fle}, and the H-VFN scheme~\cite{Kusina:2013slm}), to  
inclusive charm/bottom production in proton-proton collisions (S-ACOT-Massive-Phase-Space(-MPS) scheme~\cite{Xie:2019eoe,Xie:2021ycd,Xie:2022sqa}), and to $D$-meson hadroproduction (S-ACOT-$m_T$~\cite{Helenius:2018uul}).

The practical implementation of the ACOT scheme is facilitated by introducing the concept of \textit{subtraction}  and \textit{residual} HQ PDFs. They consist of convolutions between PDFs and universal operator matrix elements (OMEs) which represent the transition from a massless parton to an HQ and include the HQ mass dependence. Subtraction PDFs are precalculated and provided to external users as part of the CT18 family of PDFs in terms of interpolation tables to allow for fast calculations. 

For the purpose of illustration, we will use  \textit{subtraction} PDFs in the calculation of the hadronic production of a $Z$ boson in association with at least one $b$-quark jet beyond the lowest order in QCD, using theory predictions for the differential cross section obtained in previous work~\cite{FebresCordero:2008ci,FebresCordero:2009xzo,Figueroa:2018chn} by two of the authors.   
Other examples of key hadronic processes which may benefit from the implementation of a GMVFN scheme are Higgs and vector boson production in proton-proton collisions $pp\rightarrow H, \gamma^*/Z/W^{\pm}$, and heavy-flavor production in DIS (see for instance~\cite{Berge:2005rv,Belyaev:2005bs,Pietrulewicz:2017gxc,Hoang:2015iva,Gauld:2021zmq,Barontini:2024xgu}). In particular, a precise determination of DIS structure functions in PDF fits requires a GM scheme to accurately predict key scattering rates at the LHC. As more high-precision measurements from the LHC become available, it is desirable to have GMVFN schemes extended to NNLO and beyond in the case of proton-proton reactions.

The associated production of a $Z$ boson with charm- or bottom-quark jets in proton-proton collisions is of particular relevance in the study of HQ PDFs since it provides direct access to $c$ and $b$ PDFs. Indeed these processes have received great theoretical and experimental attention over the years. $Z$ + $b$-jets cross sections have been recently measured by CMS~\cite{Chatrchyan:2014dha,Khachatryan:2016iob,Sirunyan:2020lgh,CMS:2021pcj} and ATLAS~\cite{Aad:2014dvb,Aad:2020gfi,ATLAS:2024tnr} at 7, 8, and 13 TeV center of mass energies. Cross section measurements at 7 TeV for the same process in the forward region have been performed at LHCb~\cite{Aaij:2013nxa,Aaij:2014gta}, and new measurements are going to be available in the near future.
Furthermore, recent measurements of $Z$ production in association with a charm quark at 13 TeV have been obtained by the LHCb Collaboration~\cite{LHCb:2021stx}. When compared to state-of-the-art theoretical predictions these measurements can be used in global   QCD analyses of PDFs~\cite{Guzzi:2022rca,Ball:2022qks} to investigate nonperturbative charm- and bottom-quark contributions in the proton.

Theory predictions for $Z$ + $b$ jets have been calculated in the 4 flavor scheme (4FS) and in the massless and massive-$b$ 5 flavor scheme (5FS)~\cite{FebresCordero:2008ci,FebresCordero:2009xzo,Figueroa:2018chn,Forte:2019hjc,Forte:2018ovl,Krauss:2016orf,Lim:2016wjo,Bonvini:2015pxa,Bonvini:2016fgf,Forte:2015hba,Maltoni:2012pa,Caola:2011pz,Campbell:2008hh,Dawson:2003kb,Maltoni:2003pn,Krauss:2017wmx,Bagnaschi:2018dnh,deFlorian:2016spz,Gauld:2020deh,Mazzitelli:2024ura}. It has been observed that 4FS and 5FS theory predictions give compatible results and are expected to provide complementary information once they are consistently matched. A very recent calculation for the fixed-order theory prediction for $Z$ + $b$ jet at ${\cal O} (\alpha_s^3)$ in QCD~\cite{Gauld:2020deh} combines ZM-VFN at NNLO and FFN at NLO within the fixed-order-next-to-leading logarithmic (FONLL) scheme~\cite{Forte:2010ta}. The ACOT scheme results presented in this paper offer an alternative approach facilitated by the introduction of a suitably crafted set of residual PDFs.

In the following, Sec.~\ref{sec:general-framework} will review the general framework behind the application of ACOT to proton-proton collisions with the production of at least one HQ, and we will then specialize the discussion in Secs.~\ref{sec:ZQ-hard-scattering} and \ref{sec:SubResPDFs} to the case of $Z$-boson production with at least one bottom quark. 
Hadronic cross sections in an ACOT-type scheme
are discussed in Sec.~\ref{sec:Technical}.
Section~\ref{sec:ResPDFNumerical} investigates the quantitative behavior of PDF subtractions and residuals introduced in our formalism. Results for the case of $Z$-boson production with at least one $b$ jet will be presented in Sec.~\ref{sec:results}, while future applications will be discussed in our conclusive Sec.~\ref{sec:outlook}.

\section{General framework }
\label{sec:general-framework}

We begin with the familiar form of a differential cross section in the context of collinear perturbative QCD (pQCD) factorization for a generic hadronic process $A+B\rightarrow F+X$ whose final state $F$ contains at least one heavy quark $Q$ or antiquark $\bar Q$:
\begin{equation}
  \label{eq:sigma-hadronic}
  \frac{\dd\sigma(A+B\rightarrow  F+X) }{\dd{\cal X} } =
 \sum_{i,j} \int_{x_A}^1\dd\xi_A\int_{x_B}^1\dd\xi_B\,
  f_{i/A}(\xi_A,\mu) f_{j/B}(\xi_B,\mu) \frac{\dd\widehat{\sigma}(i + j\rightarrow F+X) }{ \dd{\cal X} }\,.
\end{equation}
In our case, initial-state hadrons $A$ and $B$ are protons ($A,B=p$), $\widehat \sigma$ is the hard cross section for the parton scattering process $i+ j\to F+X$, and $f_{i/A}(\xi,\mu)$ is a PDF representing the probability of finding a parton $i$ with a fraction $\xi$ of parent's momentum in $A$.  The differential cross section depends on $M_F$ ---the invariant mass of $F$, which can serve as a hard scale in the process. It may also depend on other distinct momentum scales that are collectively included in ${\cal X}$. For the collinear factorization to be applicable, we assume that $\sqrt s$, $M_F$, and other momenta in ${\cal X}$ are close in their orders of magnitude. This also implies sufficiently large Born-level momentum fractions $x_{A,B} \equiv (M_F/\sqrt{s})\exp(\pm y_F)$ in the lower limits of the integrals over $\xi_{A,B}$.  

For simplicity, $\mu$ denotes both the renormalization and factorization scales.
The parton indices $i$ and $j$ run over $N_f$ quark and antiquark species and the gluon ($i = 0$). $q$ and $Q$ denote light and heavy quarks respectively. $N_f$, the number of active quark flavors in the QCD coupling strength and PDFs at the scale $\mu$, is selected as a part of the renormalization procedure. It does not need to coincide with $N_f^{fs}$, the number of quark flavors that can be physically produced in the final state at a given scattering energy $\sqrt{s}$~\cite{Tung:2006tb}. The independence of $N_f$ from $N_f^{fs}$ opens the possibility for applying factorization schemes with different $N_f$ values, such as the FFN or VFN scheme, at the same kinematical point.

We will now remind the reader of the steps taken to derive the hard cross section $d\widehat \sigma/d\mathcal{X}$ on the right-hand side of Eq.~(\ref{eq:sigma-hadronic}) in the approach adopted in QCD factorization theorems~\cite{Collins:1998rz,Collins:1989gx,Collins:2011zzd}. We start by computing the cross section $d\sigma/d\mathcal{X}$ for the parton-level process $i+j\to F+X$. We apply UV renormalization to $d\sigma/d\mathcal{X}$ and then identify its infrared-safe part $d\widehat \sigma/d\mathcal{X}$ by factoring out parton-level PDFs $f_{i/k}$.

In more detail, we denote 
\begin{equation}
\begin{aligned}
& G_{ij} \equiv \frac{\dd\sigma(i+ j\rightarrow F+X) }{\dd{\cal X} } \mbox{ after UV renormalization},  \\
& H_{km} \equiv \frac{\dd\widehat{\sigma}(k+ m\rightarrow F+X) }{\dd{\cal X} } , 
\label{GH}
\end{aligned}
\end{equation}
and write the factorization relation in Eq. (\ref{eq:sigma-hadronic}) {\it at the parton level} as:
\begin{equation} 
\begin{aligned}
\label{eq:gij-main}
G_{ij}\left( x_A, x_B \right) &=\sum_{k,m} \int_{x_A}^1 {\dd\xi_A} \int_{x_B}^1 {\dd\xi_B}\,f_{k/i}(\xi_A)
f_{m/j}(\xi_B) H_{km}\left( \widehat{x}_A, \widehat{x}_B \right)\\
&\equiv [f_{k/i} \cA
H_{km}\cB f_{m/j}](x_A, x_B).
\end{aligned}
\end{equation}
Among all kinematic variables, we explicitly indicate dependence on collinear variables $x_i$, $\xi_i$, and $\widehat{x}_i \equiv x_i/\xi_i$. 
The right-hand side is summed over parton flavor indices $k$ and $m$. In the shorthand notation, we will generally omit the explicit sum symbol for summation over repeating flavor indices. Here and in the following we introduced convolutions, denoted by $\cA$ and $\cB$, where the triangular arrows point to the corresponding integration variable ($\xi_A$ or $\xi_B$) that is integrated in the convolution, i.e.
\begin{eqnarray}
\label{eq:cAB}
\left[f\cA H\right](x_A, x_B) \equiv \int_{x_A}^{1} \frac{\dd\xi_A}{\xi_A} f(\xi_A) H(\widehat{x}_A, x_B), \\  
\left[H\cB f\right](x_A, x_B) \equiv \int_{x_B}^{1} \frac{\dd\xi_B}{\xi_B} H(x_A,\widehat{x}_B) f(\xi_B).
\end{eqnarray}
For a convolution with one variable, we then have
\begin{eqnarray}
\label{eq:otimes}
\int_{x}^{1} \frac{\dd\xi}{\xi} f(\xi) g\left(\frac{x}{\xi}\right) = \left[f\cA g\right](x) = \left[g\cB f\right](x).
\end{eqnarray}

Given the known operator definitions for parton-level distributions $f_{k/i}$ (where both $i$ and $k$ are partons), we can solve Eq.~(\ref{eq:gij-main}) for $H_{km}$ at successive orders of pQCD. 
In the modified minimal subtraction ($\overline{\rm MS}$) scheme, the parton-level PDFs are given in terms of matrix elements of bilocal field operators that can be computed in perturbation theory. Their definition is reviewed in~\cite{Guzzi:2011ew} and can be found in the main literature~\cite{Collins:1998rz,Collins:2011zzd}. 

This solution gives us the hard-scattering cross sections $H_{ij}$ ---purely perturbative objects that are independent of the type of initial-state hadrons and collinear radiation. They can be combined with PDFs in the proton, $f_{i/A}(\xi,\mu)$, to obtain the hadron-level cross section, $\sigma(A+B\to F+X)$, in Eq.~(\ref{eq:sigma-hadronic}).

The functions $G_{ij}$, $H_{ij}$, and $f_{i/j}$ can be expanded
as a series in $a_{s}\equiv\alpha_{s}(\mu,N_{f})/(4\pi)$,  
\begin{equation}\begin{aligned}
&G_{ij}(x_A,x_B) =  G_{ij}^{(0)}(x_A,x_B)+a_{s}\, G_{ij}^{(1)}(x_A,x_B)+a_{s}^{2}\, G_{ij}^{(2)}(x_A,x_B)+\dots,
\\
&H_{ij}(\widehat{x}_A,\widehat{x}_B) =  H_{ij}^{(0)}(\widehat{x}_A,\widehat{x}_B)+a_{s}\, H_{ij}^{(1)}(\widehat{x}_A,\widehat{x}_B)+a_{s}^{2}H_{ij}^{(2)}(\widehat{x}_A,\widehat{x}_B)+\dots,
\\
&f_{i/j}(\xi) =  \delta_{ij}\delta(1-\xi)+a_{s}\, A_{ij}^{(1)}(\xi)+a_{s}^{2}A_{ij}^{(2)}(\xi)+a_{s}^{3}A_{ij}^{(3)}(\xi)+\dots\,,
\label{eq:expansions} 
\end{aligned}\end{equation}
where the $A_{ij}^{(n)}$ $(n=1,2,\dots)$ in Eq.~(\ref{eq:expansions}) are perturbative coefficients, technically calculated from OMEs
(OMEs) 
which emerge from mass factorization~\cite{Witten:1975bh,Buza:1995ie,Buza:1996wv}. 

Substituting Eq.~(\ref{eq:expansions}) into Eq.~(\ref{eq:gij-main}) and equating the coefficients of each $a_s$ order, we can solve for $H_{ij}^{(n)}$.  At leading order, we obtain
\begin{equation}
\label{eq:h0_ij}
H^{(0)}_{ij}(x_A,x_B) = G^{(0)}_{ij}(x_A,x_B),
\end{equation}
and at the next-to-leading order (assuming summation over the intermediate flavor $k$):
\begin{equation}
\label{eq:h1_ij}
H^{(1)}_{ij}(x_A,x_B)= G^{(1)}_{ij}(x_A,x_B) - [A^{(1)}_{ki} \cA H^{(0)}_{kj}](x_A,x_B)
- [H^{(0)}_{im} \cB A^{(1)}_{mj}](x_A,x_B)\,.
\end{equation}
At order $a_s^2$, we get
\begin{equation}
\label{eq:h2_ij}
\begin{aligned}
H^{(2)}_{ij}(x_A,x_B) &= G^{(2)}_{ij}(x_A,x_B)-
 [A^{(1)}_{ki} \cA H^{(1)}_{kj}](x_A,x_B) -[H^{(1)}_{im} \cB A^{(1)}_{m j}] (x_A,x_B) &
\\
&-[A^{(2)}_{ki}\cA H^{(0)}_{kj}](x_A,x_B)
-[ H^{(0)}_{im} \cB  A^{(2)}_{m j}] (x_A,x_B)& 
\\
&-[ A^{(1)}_{ki} \cA  H^{(0)}_{km} \cB  A^{(1)}_{m j}](x_A, x_B)\,,
\end{aligned}\end{equation}
while the order $a_s^3$ contribution is
\begin{equation}
\label{eq:h3_ij}
\begin{aligned}
H^{(3)}_{ij}(x_A,x_B) &= G^{(3)}_{ij}(x_A,x_B)-[A^{(1)}_{ki}\cA H^{(2)}_{kj}](x_A,x_B) - 
[H^{(2)}_{im} \cB A^{(1)}_{m j}](x_A,x_B) &
\\
&- [ A^{(2)}_{ki} \cA  H^{(1)}_{kj}](x_A,x_B)
- [ H^{(1)}_{im} \cB A^{(2)}_{m j}](x_A,x_B)& 
\\
&- [ A^{(3)}_{ki}\cA H^{(0)}_{kj}] (x_A,x_B)
- [H^{(0)}_{im} \cB A^{(3)}_{m j}] (x_A,x_B)& 
\\
& - [A^{(1)}_{ki} \cA  H^{(1)}_{km} \cB A^{(1)}_{m j} ] (x_A,x_B)\\
&  - [A^{(2)}_{ki} \cA  H^{(0)}_{km} \cB A^{(1)}_{m j} ] (x_A,x_B)
- [A^{(1)}_{ki} \cA  H^{(0)}_{km} \cB A^{(2)}_{m j} ] (x_A,x_B).
\end{aligned}\end{equation}
We notice that the various terms in the expansion of the hard-scattering function can retain the dependence on quark masses if needed. 

As discussed in Ref.~\cite{Guzzi:2011ew}, the $A^{(n)}_{ij}$ can take two different forms depending on the masses of partons $i$ and $j$. If $i$ and $j$ are massless, the $A^{(n)}_{ij}$ functions contain the DGLAP splitting functions $P^{(n)}_{ij}$ times $1/\epsilon$ poles (from dimensional regularization), plus a finite part $P^{\prime (n,l)}_{ij}$ with logs and finite terms.\footnote{
In this work, $P_{ij}^{(n)}(\xi)$ denotes the component of the DGLAP splitting function of order $a_s^n$, i.e., the pQCD expansion of the splitting function $P_{ij}(x,a_s)$ takes the form
\begin{equation*}
P_{ij}(x,a_s) = a_s P_{ij}^{(1)}(x) + a_s^2 P_{ij}^{(2)}(x) + a_s^3 P_{ij}^{(3)}(x) + \dots\, .
\end{equation*}
} 
This can be written as    
\begin{equation}\begin{aligned}
A^{(n)}_{ij}(\xi,\mu^2) = \sum_{l=1}^{n}\left(\frac{1}{\epsilon}\right)^{l} P^{(n,l)}_{ij}(\xi) +\sum_{l=0}^{n}\ln^l\left(\frac{\mu^2}{\mu^2_{\rm IR}}\right) P^{'(n,l)}_{ij}(\xi),
\end{aligned}\end{equation}
where $\mu$ is the factorization scale, and $\mu_{\rm IR}$ is the scale parameter of dimensional regularization in the infrared limit.
Here, $P_{ij}'^{(n,0)}(\xi)$ refers to nonlogarithmic terms, which can be renormalization-scheme dependent and start differing from zero for $n\geq 2$. 

On the other hand, if a massive quark with mass $m_{Q}$ is emitted from a massless parton (e.g., in $g\rightarrow Q\bar Q$ or $u \rightarrow g\rightarrow Q\bar Q$), the 
$A^{(n)}_{Qj}$ consists solely of logarithms involving the mass $m_Q$ and scale independent terms:
\begin{equation}\begin{aligned}
A^{(n)}_{Qj} \left(\xi,\frac{\mu^2}{m_Q^2}\right)= \sum_{l=0}^{n}\ln^l\left(\frac{\mu^2}{m_Q^2}\right) a_{Qj}^{(n,l)}(\xi)\,,
\label{massive-OMEs}
\end{aligned}\end{equation}
where the $a_{Qj}^{(n,l)}(\xi)$ functions can be found in~\cite{Buza:1996wv,Buza:1995ie}. For example, for the $g\rightarrow Q\bar{Q}$ splitting, $A_{Qg}^{(1)}(\xi)=2\: P_{Qg}^{(1)}(\xi)\,\ln\left(\mu^{2}/m_{Q}^{2}\right)$.
The $A^{(n)}_{Qj}$ functions are finite for $m_Q\neq 0$ but diverge when $m_Q \to 0$. For massive quark production, they appear in the subtraction terms that relate $H_{ij}^{(n)}(x_A,x_B)$ to $G_{ij}^{(n)}(x_A,x_B)$. 
The $A^{(n)}_{Qj}$ functions for the highest available values of $n$ can be found in Refs.~\cite{Ablinger:2024qxg,Ablinger:2024xtt,Ablinger:2023ahe,Bierenbaum:2022biv,Ablinger:2022wbb,Behring:2021asx,Ablinger:2020snj,Ablinger:2019etw,Ablinger:2018brx,Ablinger:2017xml,Ablinger:2017err,Ablinger:2014tla,Ablinger:2014uka,Behring:2014eya,transitionAqg,Ablinger:2012qm,Blumlein:2012vq,Blumlein:2011mi,Ablinger:2010ty,Bierenbaum:2009zt,Bierenbaum:2008yu,Bierenbaum:2007qe,Bierenbaum:2007dm}.

In this article, we denote infrared-safe contributions by a caret symbol. 
In the next section, we derive the hard cross sections $H^{(n)}_{ij}$ (i.e., $\widehat \sigma^{(n)}(i+j\to F+X)$) for channels with massive quarks according to Eqs.~(\ref{eq:h0_ij}--\ref{eq:h3_ij}). As the first step, we apply dimensional regularization for initial-state massless partons and subtract the associated $1/\epsilon$ singularities from $G_{ij}^{(n)}$, which produces the infrared-safe parts (with respect to massless partons) of $G_{ij}^{(n)}$ given by 
\begin{equation}\begin{aligned}
\widehat{G}^{(n\leq k)}_{ij}\left(x_A,x_B,\frac{Q^2}{\mu^2}\right) 
&=G^{(n\leq k)}_{ij}\left(x_A,x_B,\frac{Q^2}{\mu^2_{\rm IR}}, \frac{1}{\epsilon}\right) 
\\
&-\sum_{p=0}^{n\leq k}\sum_{l=0}^{n-p} [A^{(n-p-l)}_{ki}\cA  H^{(p)}_{k m} \cB  A^{(l)}_{m j}]  (x_A,x_B)\,
\label{general-infrared-safe}
\end{aligned}\end{equation}
for $n$ up to order $k$.
Here, $A_{ij}^{(0)}$ only contains a Kronecker delta and a $\delta$-function according to the last line in Eq.~(\ref{eq:expansions}). $\widehat G_{ij}$ is finite even if $G^{(n)}_{ij}$ and the PDF coefficients $A_{km}^{(p)}$ of massless flavors contain $1/\epsilon^p$ poles with $p>0$.

\section{Application to $ZQ$ associated production}
We will now apply the prescription described in Sec.~\ref{sec:general-framework} to $ZQ$ associated production at hadron colliders. We start by categorizing the hard-scattering contributions into flavor creation (FC) and flavor excitation (FE) types in Sec.~\ref{sec:ZQ-hard-scattering}. We then show that the prescription of Sec.~\ref{sec:general-framework} appropriately subtracts the overlapping part between the FC and FE to avoid the double counting. To facilitate this subtraction in a practical computation, Sec.~\ref{sec:SubResPDFs} introduces subtraction and residual heavy-quark PDFs. Finally, Sec.~\ref{sec:Technical} presents a master formula for the hadronic cross section in terms of the residual PDFs.

\begin{figure}[tb]
\includegraphics[width=0.24\textwidth]{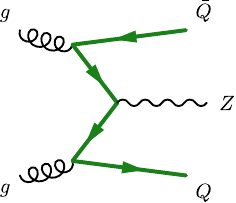}
\includegraphics[width=0.24\textwidth]{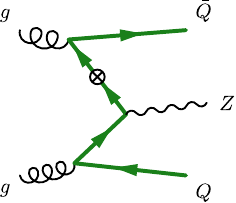}
\includegraphics[width=0.24\textwidth]{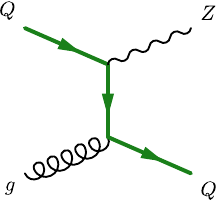}
\caption{Representative Feynman diagrams for $Z+Q$ production at the lowest order in QCD. The diagrams on the left and on the right represent the FC and FE contributions, respectively. The middle diagram represents a subtraction term. The $\otimes$ symbol on the quark line represents a collinear splitting.}
\label{fig:VFN@LO}
\end{figure}

\subsection{Hard-scattering functions}
\label{sec:ZQ-hard-scattering}
In the production of the final state $F=Z+Q$, the flavor indices $i$ and $j$ in the partonic scattering channels $i+j\to Z+Q+X$ run over light (anti)quarks $q$, the heavy (anti)quark $Q$ (charm or bottom), and gluons, while $X$ denotes inclusively other constituents of the final state.

It will be helpful to categorize the Feynman diagrams for parton-scattering cross sections $G_{ij}$ into FC terms (see, {\it e.g.}, the first diagram in Fig.~\ref{fig:VFN@LO}) and FE terms (see, {\it e.g.}, the third diagram in Fig.~\ref{fig:VFN@LO}). The FC terms contain HQ only in the final state, and are present both in 
the FFN as well as in any VFN scheme, including the ACOT scheme adopted in this paper. We always calculate these contributions with $m_Q\neq 0$. 

On the other hand, the FE terms correspond to HQ-initiated processes, where the large collinear logarithms have been conveniently resummed into an HQ PDF. We focus on the case when the HQ PDF is perturbatively generated from light-parton contributions, at the first order through the gluon splitting ($g\rightarrow Q\bar{Q}$). In the hadronic cross section, we will ascribe an extra power of $\alpha_s$ to each HQ PDF appearing 
in the FE initial states.\footnote{Generally, there is also a possibility that the heavy-quark PDF is generated at low-scale $Q_0$ from a higher-twist term \cite{Brodsky:1984nx,Hou:2017khm}, giving rise to the intrinsic heavy-quark contribution \cite{Brodsky:1980pb}. In this case, we still expect the relative suppression of the HQ PDF compared to the light sea PDFs, arising from the overall factor of order $\alpha_s^2 \Lambda_{\rm QCD}^2/m_Q^2$ in the generating higher-twist term \cite{Hou:2017khm}.} 
Furthermore, the $m_Q$ dependence of FE cross sections and the corresponding subtractions with the same hard functions cancels up to the higher order in $a_s$ and power-suppressed terms~\cite{Hou:2017khm}. We further discuss this point in Sec.~\ref{sec:Technical}. 
The FE cross sections are calculated with $m_Q=0$ in the S-ACOT scheme and with $m_Q\neq 0$ in the ACOT scheme. In Sec.~\ref{sec:results}, we will consider both cases. 

The lowest-order hard function $H_{ij}$ is of order $a_s$ and proceeds through the $Qg$ channel (FE, 
which we explicitly indicate by the corresponding function $G_{ij}$).\footnote{Here and in the following it is understood that counterpart external states with antiquarks 
($\bar{Q}$, $\bar{q}$) instead of quarks ($Q$, $q$) are also to be considered.} As a consequence $H^{(0)}_{ij}=G^{(0)}_{ij}=0$, and Eq.~(\ref{eq:h1_ij}) reduces to
\begin{equation}\begin{aligned}
H^{(1)}_{Qg}(x_A,x_B)=\left. G^{(1)}_{Qg}(x_A,x_B)\right|_{\rm FE}\,.
\label{eq:H1-Qg}
\end{aligned}\end{equation}

At the next order, Eq.~(\ref{eq:h2_ij}) gives $H^{(2)}_{ij}$ in terms of the lowest-order FC contribution, $\widehat{G}^{(2)}_{ij}$, from which terms proportional to $H^{(1)}_{ij}$ in the collinear limit have been subtracted, namely:
\begin{equation}
\begin{aligned}
H^{(2)}_{ij}(x_A,x_B) &= \widehat{G}^{(2)}_{ij}(x_A,x_B) -  \left[A^{(1)}_{ki} \cA H^{(1)}_{kj}\right](x_A,x_B) 
-  \left[ H^{(1)}_{im} \cB A^{(1)}_{mj} \right](x_A,x_B)\, .
\label{eq:H2-ZQ}
\end{aligned}
\end{equation}
Two different combinations of the initial-state partons $i$ and $j$, i.e., $gg$ and $q\bar{q}$ (with light quarks), may occur at this order, with only the $gg$ channel generating subtraction terms due to the $g\rightarrow Q\bar{Q}$ collinear configuration:
\begin{equation}
H^{(2)}_{gg}(x_A,x_B)  = \left.G^{(2)}_{gg}(x_A,x_B)\right|_{\rm FC} -  \left[A^{(1)}_{Qg} \cA H^{(1)}_{Qg}\right](x_A,x_B) - \left[H^{(1)}_{gQ} \cB A^{(1)}_{Qg} \right](x_A,x_B)\,.
\label{eq:H2-gg}
\end{equation}
For the $q\bar{q}$ channel (not shown in Fig.~\ref{fig:VFN@LO}), Eq.~(\ref{eq:H2-ZQ}) reduces to 
\begin{equation}
H^{(2)}_{q\bar q}(x_A,x_B)  = \left.G^{(2)}_{q\bar q}(x_A,x_B)\right|_{\rm FC}\,.
\label{eq:H2-qq}
\end{equation}
This channel does not contain initial-state $g\to Q\bar Q$ splittings, but it does contain finite-state ones that are further discussed in Sec.~\ref{sec:results}. Being the FC contributions, both $G^{(2)}_{gg}$ and $G^{(2)}_{q\bar q}$ have the exact kinematical dependence on $m_Q$. In contrast, the $G^{(2)}_{Q\bar Q}$ initiated by heavy quarks is an FE contribution and technically contributes at order $a_s^4$ due to the additional suppression of the heavy-quark PDF.
Up to order $a_s^2$, the hard-scattering function for $Z+Q$ production is then given by
\begin{equation}\begin{aligned}
a_s H^{(1)} + a_s^2 H^{(2)} & = a_s H^{(1)}_{Qg}(x_A,x_B ) + a_s^2 H^{(2)}_{gg}(x_A,x_B )
+ a_s^2 H^{(2)}_{q\bar q}(x_A,x_B ),
\label{eq:H1+H2}
\end{aligned}\end{equation}
where $H^{(1)}_{Qg}$  is given in Eq.~(\ref{eq:H1-Qg}), while
$H^{(2)}_{gg}$ and $H^{(2)}_{q\bar{q}}$ are given in Eqs.~(\ref{eq:H2-gg})-(\ref{eq:H2-qq}).

\begin{figure}
\includegraphics[width=0.24\textwidth]{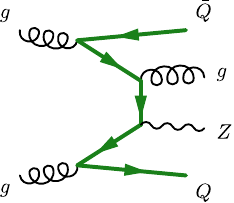}
\includegraphics[width=0.24\textwidth]{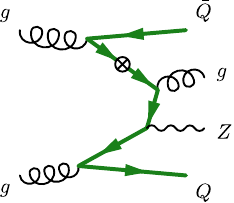}
\includegraphics[width=0.24\textwidth]{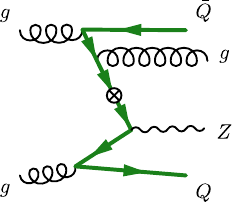}
\includegraphics[width=0.24\textwidth]{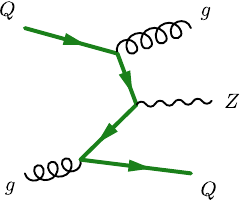}
\caption{Representative Feynman diagrams at the next-to-leading order in the $gg$ channel. FC and FE terms are represented by the leftmost and rightmost diagrams, respectively. The subtraction terms are represented by the two diagrams in the middle. Virtual contributions are not displayed here, but are included in the calculation. }
\label{fig:VFNgg}
\end{figure}
\begin{figure}
\centering
\includegraphics[width=0.24\textwidth]{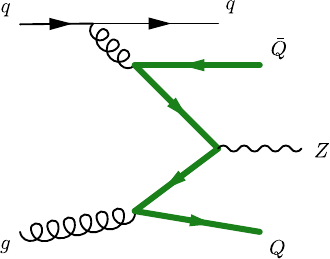}
\includegraphics[width=0.24\textwidth]{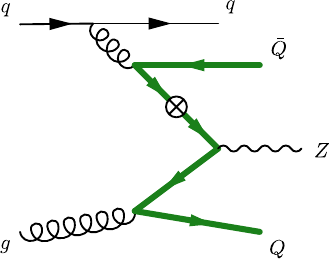}
\includegraphics[width=0.24\textwidth]{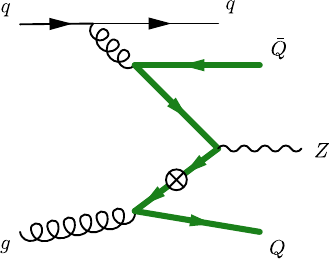}
\includegraphics[width=0.24\textwidth]{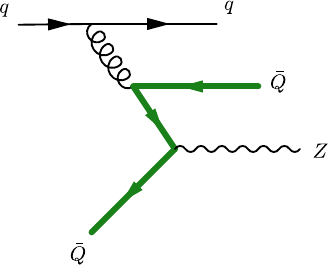}
\caption{Same as in Fig.~\ref{fig:VFNgg}, but for the $qg$ channel.}
\label{fig:VFNqg}
\end{figure}

Once embedded in Eq.~(\ref{eq:sigma-hadronic}), the first two terms in Eq.~(\ref{eq:H1+H2}) provide an infrared-safe combination of the $gg$ and $Qg$ channels at order $a_s^2$ according to the ACOT-type scheme.  
Figure~\ref{fig:VFN@LO} depicts the representative diagrams at this perturbative order,  where the diagram on the left belongs to the FC class, the one on the right to the FE class, and the  middle diagram represents the subtraction terms in Eq.~(\ref{eq:H2-gg}). The $\otimes$ symbol indicates a collinear splitting in the subtraction term.

Proceeding now to the ${\cal O}(a_s^3)$ contributions, we apply the master equation (\ref{eq:h3_ij}) for $H^{(3)}_{ij}$ to the respective contributions, examples of which are illustrated in Fig.~\ref{fig:VFNgg} for the $gg$ channel and in Fig.~\ref{fig:VFNqg} for the $qg$ channel. As in Fig.~\ref{fig:VFN@LO}, the FC and FE terms are represented by the diagrams on the far left and far right, respectively, while the 
subtraction terms are represented by the two diagrams in the middle. In addition to the shown diagrams, our calculation also includes virtual contributions as well as the 
$q\bar{q}\to ZQ\bar{Q}$ annihilation at this order, with the latter one being free of the logarithmic enhancement giving rise to the resummed heavy-flavor PDFs and therefore not associated with an HQ subtraction term. We also include the respective FE hard functions in the $Qg$ and $Qq$ channels.

Explicitly, the hard functions contributing at order $a_s^3$ to the cross section 
are given by
\begin{align}
H^{(2)}_{Qi}(x_A,x_B) &=\left.\widehat{G}^{(2)}_{Qi}(x_A,x_B)\right|_{\rm FE} \hspace{10pt} \mbox{ for }i=g,q,\bar{q}\,, 
    \label{eq:H2-Qi} \\
H^{(3)}_{ij}(x_A,x_B) &= \left.\widehat{G}^{(3)}_{ij}(x_A,x_B)\right|_{\rm FC} -[A^{(1)}_{Qi}\cA H^{(2)}_{Qj}](x_A,x_B)  - 
[H^{(2)}_{iQ} \cB A^{(1)}_{Q j}](x_A,x_B)  \nonumber \\
&- [ A^{(2)}_{Qi} \cA  H^{(1)}_{Qj}](x_A,x_B) 
- [ H^{(1)}_{iQ} \cB A^{(2)}_{Q j}](x_A,x_B) \hspace{10pt} \mbox{ for }i,j=g,q,\bar{q}\,,
\label{eq:H3-ij} \\
H^{(3)}_{q\bar{q}}(x_A,x_B)&=\left.\widehat{G}^{(3)}_{q\bar{q}}(x_A,x_B)\right|_{\rm FC}\,.
    \label{eq:H3-qq}
\end{align}
Note that, for $H_{gg}^{(3)}$ in Eq.~(\ref{eq:H3-ij}), $\widehat{G}_{gg}^{(3)}$ already accounts for the subtraction of the $1/\epsilon$ poles, including $A^{(1)}_{Qg} \cA H^{(1)}_{Qg} \cB A^{(1)}_{g g}$
(see Eq.~(\ref{general-infrared-safe})).
The result for $H_{ij}$ up to ${\cal O}(a_s^3)$ is then given by 
\begin{equation}\begin{aligned}
a_s H^{(1)}+a_s^2 H^{(2)}+a_s^3 H^{(3)} 
&=     
a_s H^{(1)}_{Qg}(x_A,x_B) 
+a_s^2 H^{(2)}_{gg}(x_A,x_B)
+a_s^2 H^{(2)}_{q\bar q}(x_A,x_B)
\\
&+a_s^2 H^{(2)}_{Qg}(x_A,x_B) 
+a_s^2 H^{(2)}_{Qq}(x_A,x_B) 
\\
&+ a_s^3 H^{(3)}_{gg}(x_A,x_B)
+ a_s^3 H^{(3)}_{qg}(x_A,x_B)
+ a_s^3 H^{(3)}_{q\bar q}(x_A,x_B) \,.
\end{aligned}
\label{H-matching-as3}
\end{equation}

\subsection{NLO hadronic cross sections in terms of PDF subtractions and residuals} 
\label{sec:SubResPDFs}

Once the hard-scattering function $H^{(n)}_{ij}$ coefficients are determined, they are convoluted with the initial-state proton PDFs and summed over the various channels to obtain the hadronic cross section in Eq.~(\ref{eq:sigma-hadronic}), 
which, using the notation introduced in Eq.~(\ref{eq:cAB}) and dropping ``$d{\cal X}$'' for brevity, can be schematically written as 
\begin{equation}\begin{aligned}
\dd\sigma =
\sum_{i,j}f_{i/A} \cA
 \left[a_s H^{(1)}+a_s^2 H^{(2)} + a_s^3 H^{(3)} +\dots\right]_{ij}
 \cB f_{j/B}\,.
\end{aligned}\end{equation}
By expanding the expressions for $H^{(n)}_{ij}$, the various subtraction (``sub'') terms up to $O(a_s^3)$ can be collected as follows:  
\begin{equation}\begin{aligned}
-\dd\sigma_{\rm sub} = 
&-a_s^2 \left[ g \cA A^{(1)}_{Qg} \cA H^{(1)}_{Qg}\right] \cB g
- a_s^3 \left[\sum_{i, j =g,q,\bar q} f_i \cA A^{(2)}_{Qi} \cA H^{(1)}_{Qj} \right] \cB f_j 
\\
& - a_s^3 \left[\sum_{i, j =g,q,\bar q} f_i \cA A^{(1)}_{Qi} \cA H^{(2)}_{Qj} \right] \cB f_j 
+ ({\rm exch.})
\,,
\label{eq:Xsec-sub}
\end{aligned}\end{equation}
where the summation over gluons and active (anti)quarks (generically indicated as $q$ and $\bar q$) is understood. ``$({\rm exch.})$'' indicates that one must include terms with exchanged $x_A$ and $x_B$ convolutions in addition to the shown ones. Notice that the terms grouped within big square brackets in Eq.~(\ref{eq:Xsec-sub}) are the first terms in the expansion of the HQ PDF ($f_Q$). One can then introduce an HQ PDF \textit{subtraction} $\tilde{f}_Q$ whose first two orders in $a_s$ are identified as~\cite{Xie:2019eoe,Xie:2021ycd}
\begin{equation}
\begin{aligned}
&\tilde{f}_Q^{(1)} = a_s [A^{(1)}_{Qg} \cB g],~
\tilde{f}^{(2)}_Q = a_s^2 \sum_{i=g, q,\bar q} [A^{(2)}_{Qi} \cB f_i].
\end{aligned}
\label{eq:sub-PDFs-def}
\end{equation}
Equation~(\ref{eq:Xsec-sub}) now reads as\footnote{At this perturbative order $H^{(1)}_{Qq}=0$ and $A^{(1)}_{Qq}=0$ (same for $q\rightarrow \bar q$).}
\begin{equation}
\label{eq:Xsec-sub-subPDF}
\begin{aligned}
-\dd\sigma_{\rm sub}
=&-a_s\, \tilde{f}_{Q}^{\rm(NLO)}\cA H^{(1)}_{Qg}\cB g
- a_s^2\, \sum_{i=g, q,\bar q}  \tilde{f}_Q^{(1)}\cA  H^{(2)}_{Qi}\cB f_i  \, 
\end{aligned}\end{equation}
in terms of the NLO PDF subtraction, i.e.,
\begin{equation}\begin{aligned}
\tilde{f}_Q^{\rm(NLO)}(x,\mu) \equiv \tilde{f}_Q^{(1)}+ \tilde{f}_Q^{(2)}=a_s [ A^{(1)}_{Qg} \cB g](x,\mu) + a_s^2 \sum_{i=g,q,\bar q} [A^{(2)}_{Qi} \cB f_i](x,\mu)\,,  
\label{eq:sub-PDFs-NLO}
\end{aligned}\end{equation}
where ``NLO'' refers to the $a_s$ order of the subtraction coefficients. 

The subtraction terms in Eq.~(\ref{eq:Xsec-sub-subPDF}) can be grouped together with the FE contributions in Eq.~(\ref{H-matching-as3}), as they share the same hard-scattering functions  which are convoluted with the corresponding HQ PDFs.  
It is therefore natural to also introduce PDF \emph{residuals}, labeled as $\delta f_Q$ and defined as the difference between the $f_Q$ and $\tilde{f}_Q$~\cite{Xie:2019eoe,Xie:2021ycd},
\begin{equation}\begin{aligned}
\delta f_Q^{(1)} = f_{Q}  - \tilde{f}_Q^{(1)}\,,~~~~ \delta f_Q^{\rm(NLO)} = f_{Q}  - \tilde{f}_Q^{\rm(NLO)}\,.
\label{eq:res-PDFs-def}
\end{aligned}\end{equation}
As a result, the difference of the FE contributions in Eq.~(\ref{H-matching-as3}) and subtracted terms in Eq.~(\ref{eq:Xsec-sub-subPDF}) can be written in a compact form as 
\begin{equation}
\label{eq:Xsec-sub-resPDF}
\begin{aligned}
\dd\sigma_{\rm FE} - \dd\sigma_{\rm sub}
=&a_s(f_Q-\tilde{f}^{\rm{(NLO)}}_Q) \cA H^{(1)}_{Qg}\cB g\\
&+a_s^2(f_Q-\tilde{f}_Q^{(1)})\cA \left[H^{(2)}_{Qg}\cB g
+ \sum_{i=q,\bar q} H^{(2)}_{Qq} \cB f_i \right]+ ({\rm exch.})
\\
=&a_s\, \delta f_{Q}^{\rm(NLO)}\cA H^{(1)}_{Qg}\cB g
+ a_s^2\, \delta f_Q^{(1)}\cA \left[H^{(2)}_{Qg} \cB g +\sum_{i=q,\bar q} H^{(2)}_{Qq}\cB f_i \right]+({\rm exch.}).
\end{aligned}\end{equation}

We are now ready to derive the $Z+Q$ inclusive cross section in an ACOT-type GMVFN scheme by combining the FC contributions in  Sec.~\ref{sec:ZQ-hard-scattering} with the FE and subtraction contributions in terms of PDF residuals. In the abridged notation, the general structure of the cross section in such a GMVFN scheme reads
\begin{equation}\label{eq:VFN}
\dd\sigma_{\rm GMVFN}=\dd\sigma_{\rm FC}+\dd\sigma_{\rm FE}-\dd\sigma_{\rm Sub}.
\end{equation}
At NLO, the sum of FE and subtraction terms is provided in Eq.~(\ref{eq:Xsec-sub-resPDF}).
Restoring the notation $\dd\widehat{\sigma}_{ij}^{(n)}$ for the hard cross sections $H_{ij}^{(n)}$ of order $a_s^n$ (computed in Sec.~\ref{sec:ZQ-hard-scattering}) and explicitly listing the contributing partonic channels, in terms of PDF \textit{subtractions} the components of $\dd\sigma_{\rm GMVFN}^{\rm NLO}$ are
\begin{equation}
\begin{aligned}
\dd\sigma_{\rm FC}^{\rm NLO} &=
f_{g} \cA \left[a_s^2 ~\dd\widehat{\sigma}^{(2)}_{gg\to ZQ\bar{Q}} + a_s^3 ~\dd\widehat{\sigma}^{(3)}_{gg\to ZQ\bar{Q}(g)}  \right]\cB f_{g}
\\
&+\sum_{i=q, \bar q}f_{i}\cA\left[a_s^2~\dd\widehat{\sigma}^{(2)}_{q\bar{q}\to ZQ\bar{Q}}+ a_s^3~\dd\widehat{\sigma}^{(3)}_{q\bar{q}\to ZQ\bar{Q}(g)}\right] \cB f_{\bar i} \\
&+a_s^3\sum_{i=q,\bar q} \left[  f_{g} \cA  \dd\widehat{\sigma}^{(3)}_{gq\to ZQ\bar{Q}(q)} \cB f_{i} 
+ f_{i} \cA \dd\widehat{\sigma}^{(3)}_{qg\to ZQ\bar{Q}(q)} \cB f_{g}\right],
\label{eq:tot-FC}
\end{aligned}\end{equation}
\begin{equation}\begin{aligned}
\dd\sigma_{\rm FE}^{\rm NLO}
&=f_{g}\cA  \left[a_s~\dd\widehat{\sigma}^{(1)}_{gQ\to ZQ}+
a_s^2~\dd \widehat{\sigma}^{(2)}_{gQ\to ZQ(g)}\right]\cB f_{Q}
\\
&+ a_s^2 \sum_{i=q,\bar q} f_i \cA \dd\widehat{\sigma}^{(2)}_{qQ\to ZQ(q)}\cB f_{Q}
+(\rm{exch.}),
\label{eq:tot-FE}
\end{aligned}\end{equation}
\begin{equation}\begin{aligned}
\dd\sigma_{\rm sub}^{\rm NLO}
&=a_s~f_g\cA \dd\widehat{\sigma}^{(1)}_{gQ\to ZQ}\cB \tilde{f}_Q^{(\rm NLO)}
+a_s^2 ~f_g \cA 
\dd\widehat{\sigma}^{(2)}_{gQ\to ZQ(g)}\cB \tilde{f}_Q^{(1)}
\\
&+ a_s^2 \sum_{i=q,\bar q} f_i \cA  
\dd\widehat{\sigma}^{(2)}_{qQ\to ZQ(q)}\cB \tilde f_Q^{(1)}
+(\rm{exch.})\,,
\label{eq:tot-subtraction-pdf}
\end{aligned}\end{equation}
or, equivalently with the $\dd\sigma_{\rm FE}-\dd\sigma_{\rm sub}$ reorganized in terms of HQ PDF \textit{residuals}:
\begin{equation}\begin{aligned}
\dd\sigma_{\rm GMVFN}^{\rm NLO}
&= d\sigma_{\rm FC}^{\rm NLO} + a_s~f_{g}\cA\left[\dd\widehat{\sigma}^{(1)}_{gQ\to ZQ}\right]\cB \delta f_{Q}^{(\mathrm{NLO})} 
\\
&+ a_s^2~f_{g} \cA \left[ \dd\widehat{\sigma}^{(2)}_{gQ\to ZQ(g)}\right]\cB \delta f_{Q}^{(1)}
+a_s^2~\sum_{i=q,\bar q} f_i \cA \left[\dd\widehat{\sigma}^{(2)}_{qQ\to ZQ(q)}\right]\cB \delta{f}^{(1)}_Q
+(\rm{exch.})\,.
\end{aligned}
\label{eq:tot-residual-pdf}
\end{equation}
We notice that the quark-antiquark annihilation channel $q\bar{q}\to ZQ\bar{Q}$ does not contain FE terms and therefore does not involve subtraction contributions.  

\subsection{Hadronic cross sections in an ACOT-type scheme \label{sec:Technical}}
The generic GMVFN hadronic cross section obtained in Eq. (\ref{eq:tot-residual-pdf}) can be called the ``lowest mandatory order" (LMO) representation at NLO in that it retains only the unambiguous terms up to order $a_s^3$ required by order-by-order factorization and scale invariance. Any ACOT-like scheme must contain such terms. In addition, one generally can augment Eq.~(\ref{eq:tot-residual-pdf}) with extra radiative contributions from higher orders with the goal to improve consistency with the specific GMVFN scheme adopted in the fit of the used PDFs. 

The GMVFN scheme assumed for determination of CTEQ-TEA PDFs \cite{Guzzi:2011ew} with up to five active flavors is closely matched with the following additional choices:
\begin{enumerate}[label=\alph*)]
\item In Eq.~(\ref{eq:tot-residual-pdf}), we evolve $\alpha_s(\mu)$ and PDFs $f_i(\xi,\mu)$ with $N_f=5$ at $\mu \geq m_b$. The hard cross sections are also evaluated with $N_f=5$ in virtual loops both for massive and massless channels. If the virtual contributions are obtained in the $N_f=4$ scheme, they should be converted to the $N_f=5$ scheme by adding known terms to the hard cross sections \cite{Forte:2010ta,Guzzi:2011ew,Buza:1996wv,Chetyrkin:2000yt}.
\item  The sums over initial-state light quarks and antiquarks in Eq.~(\ref{eq:tot-residual-pdf}) are extended to also include the $b$-quark PDFs via the introduction of the singlet PDF $\Sigma \equiv \sum_{i=1}^{5} (f_i + \bar f_{i})$. The initial-state $b$-quark lines in the Feynman graphs can be treated on the same footing as the light-quark ones, i.e.,  the $b$-quark mass is to be omitted (retained) on the $b$-quark lines attached directly (attached through gluons) to the initial-state hadrons.\footnote{\small We already noted that the  $f_Q \cA d\widehat{\sigma}^{(2)}_{Q\bar Q\to Q\bar Q}\cB f_Q$ cross section is associated with order $a_s^4$, yet it can be included already at NLO due to its initial-state logarithmic enhancement. Compared to the $q\bar q \to Q\bar Q$ channel with different initial-state and final-state quarks, the $Q\bar Q\to Q\bar Q$  cross section includes extra contributions with $t$ and $u$ channels. In the following numerical computations, its respective contribution to $Zb$ production is not significant.}
\item Without sacrificing the ${\cal O}(a_s^3)$ accuracy, we can further replace $\tilde f^{(1)}_Q$ in Eq.~(\ref{eq:tot-subtraction-pdf}) and $\delta \tilde f^{(1)}_Q$ in Eq.~(\ref{eq:tot-residual-pdf}) by $\tilde f^{\rm (NLO)}_Q$ and $\delta \tilde f^{\rm (NLO)}_Q$, respectively.
\item The $\alpha_s$ and PDFs must be evolved at least at NLO, although evolution at NNLO is acceptable or even desirable in some contexts. In this regard, we notice that $Z+b$ production contributes to data for inclusive $pp\to ZX$ at scales $Q^2 \sim M_Z^2 \gg m_b^2$ to which the PDFs are fitted. At this $Q^2$, the third diagram $b+g\to Z + b$ in Fig.~\ref{fig:VFN@LO} contributes at ${\cal O}(\alpha_s)$ to the inclusive $Z$ hard cross section, while the difference of the first and second terms contributes at ${\cal O}(\alpha_s^2)$. Therefore, at this order, while the LMO prescription would use the leading-order (LO) PDFs for the stand-alone process, better consistency with inclusive $Z$ production is reached by using NLO PDFs with LO subtraction. When also including the diagrams of the next order in Figs.~\ref{fig:VFNgg} and \ref{fig:VFNqg}, this rule allows one to use NNLO PDFs with an NLO subtraction, and so on.\footnote{\small This freedom with choosing the PDF order in $Zb$ production is analogous to the one  in inclusive deep-inelastic scattering, for which the lowest-order Bjorken scaling allows scale-independent PDFs, while in practice one uses one higher $\alpha_s$ order in the PDF evolution, i.e., the LO PDFs evolved according to the one-loop DGLAP equation to also capture the scaling violations in DIS cross sections.} 
\item In the hard cross sections inside $\dd\sigma_{\rm FE}-\dd\sigma_{\rm sub}$, dependence on the HQ mass can be  eliminated altogether \cite{Collins:1998rz,Kramer:2000hn} or simplified \cite{Tung:2001mv},  producing a difference only in higher-order terms. The choice of the HQ mass treatment in $\dd\sigma_{\rm FE}-\dd\sigma_{\rm sub}$ matters little in $Z+b$ production and other processes in which momentum virtualities in hard-scattering cross sections are much larger than the HQ mass. On the other hand, in processes with low virtualities comparable to $m_Q$, such as DIS or $b$-quark hadroproduction, an optimized scheme such as S-ACOT-MPS leads to better perturbative convergence. Section~\ref{sec:results} will explore the numerical impact of these choices.
Different HQ mass treatments in $\dd\sigma_{\rm FE}-\dd\sigma_{\rm sub}$ may affect the cross section near the threshold. To further elaborate, in the $Q^2\gg m_Q^2$ limit, the FC $G^{(n)}_{ij}$ terms can be represented by the sum of three contributions 
\begin{equation}\begin{aligned}
\left. G_{ij}^{(n)}(x_A,x_B)\right|_{\textrm{FC}} = \left.G_{ij}^{(n)}(x_A,x_B)\right|_{m_Q=0} + \left.G_{ij}^{(n)}(x_A,x_B)\right|_{\ln({m_Q^2/Q^2})}+  \left.G_{ij}^{(n)}(x_A,x_B)\right|_{m_Q^2/Q^2}
\label{G-decomp}
\end{aligned}\end{equation}
where the first term only depends on $N_f$, the second contains the logarithmic behavior, and the third represents all power suppressed contributions $\mathcal{O}\left(m_Q^2/Q^2\right)$ \cite{Buza:1995ie,Buza:1996wv,Gauld:2021zmq}.
According to Eq.~(\ref{massive-OMEs}), the structure of the logarithmic terms can be written as follows~\cite{Buza:1995ie}:
\begin{equation}
\left.G_{ij}^{(n)}\left(x_A,x_B,\frac{Q^2}{m_Q^2},\frac{m_Q^2}{\mu^2}\right)\right|_{\ln({m_Q^2/Q^2})}=
\sum_{l=0}^{n} a_{ij}^{(n,l)}\left(x_A,x_B,\frac{m^2_Q}{\mu^2}\right) \ln^l\left(\frac{Q^2}{m_Q^2}\right). 
\end{equation}
The structure of the power-suppressed contributions can be inferred from the computation of the $H^{(n)}_{ij}$ hard functions in Eqs.~(\ref{eq:h1_ij})-(\ref{eq:h3_ij}), where collinear subtractions are performed:
\begin{equation}
\left.G_{ij}^{(n)}(x_A,x_B)\right|_{m_Q^2/Q^2}=\sum_{j=1}^{\infty} \sum_{l=1}^{n}\left[a_j \ln^l\left(\frac{m_Q^2}{Q^2}\right)+b_j\right]\left(\frac{m_Q^2}{Q^2}\right)^j, \end{equation}
where $n$ matches the power of $a_s$ in the calculation. When the FC contributions are combined with the subtraction and FE terms to obtain the cross section, the FE contributions reduce to uniquely defined expressions in the $\overline{\textrm{MS}}$ scheme when $m_Q$ is small. However, near the threshold, they may be approximated in different ways by retaining or dropping powerlike contributions $(m^2_Q/Q^2)^j$
with $j > 0$. Within the ACOT framework, one may adopt several conventions to include these powerlike contributions in a way compatible with the QCD factorization theorem~\cite{Kramer:2000hn,Tung:2001mv,Guzzi:2011ew,Xie:2019eoe}.
\end{enumerate}

Based on these considerations, it is possible to revise the GMVFN calculation as follows. In our $N_f=5$ scheme with the two-loop (non)singlet flavor combinations introduced as in Refs.~\cite{Buza:1996wv,Buza:1995ie}, where the two-loop OMEs were computed, Eq.~(\ref{eq:sub-PDFs-def}) for the subtraction PDFs is recast as
\begin{equation}\begin{aligned}
\tilde{f}_Q^{(1)} = a_s \left[ A^{S,(1)}_{Qg} \cB g\right], \quad
\tilde{f}^{(2)}_Q = a_s^2 \left[ A^{\rm PS,(2)}_{Qq} \cB \Sigma + A^{S,(2)}_{Qg} \cB g \right]\, ,
\label{sub-with-OMEs2}
\end{aligned}\end{equation}
where $\Sigma$ is the singlet PDF combination, and $A^{S,(n)}_{ij}$ and $A^{\rm PS,(n)}_{ij}$ are the singlet and pure-singlet OMEs, respectively.  The subtraction cross section in Eq.~(\ref{eq:tot-subtraction-pdf})  takes  the form
\begin{equation}\begin{aligned}
\dd\sigma_{\rm sub}^{\rm NLO}
&= \Biggl(a_s~f_g\cA \dd\widehat{\sigma}^{(1)}_{gQ\to ZQ}
+a_s^2 ~f_g \cA 
\dd\widehat{\sigma}^{(2)}_{gQ\to ZQ(g)}
\\
&+ a_s^2~ \Sigma \cA  
\dd\widehat{\sigma}^{(2)}_{qQ\to ZQ(q)}\Biggr) \cB \tilde f_Q^{\rm (NLO)}
+(\rm{exch.})\,,
\label{eq:tot-subtraction-pdf2}
\end{aligned}\end{equation}
or, if performing the expansion in terms of the OMEs up to $a_s^3$ , 
\begin{equation}
\begin{aligned}
\dd\sigma_{\rm sub}^{\rm (NLO)} &= a_s^2  g \cA \dd\widehat{\sigma}^{(1)}_{gQ\to ZQ} \cB \left[A^{S,(1)}_{Qg}\cB g\right] \\
&+ a_s^3 \Biggl\{
\left(g \cA \dd\widehat{\sigma}^{(2)}_{gQ\to ZQ} + \Sigma   \cA \dd\widehat{\sigma}^{(2)}_{qQ\to ZQ} \right) \cB \left[A^{S,(1)}_{Qg} \cB g \right]
\\
&+ g \cA \dd\widehat{\sigma}^{(1)}_{gQ\to ZQ} \cB \left[A^{\rm PS,(2)}_{Qq} \cB \Sigma   + A^{S,(2)}_{Qg} \cB g \right] \Biggr\}
+ \mbox{(exch.)}.
\end{aligned}
\label{eq:tot-subtraction-pdf3}
\end{equation}
The master equation for Eq.~(\ref{eq:tot-residual-pdf}) for the NLO cross section in the ACOT-type $N_f=5$ scheme using the residual PDF now reads as
\begin{equation}\begin{aligned}
\dd\sigma_{\rm ACOT}^{\rm NLO}
&= \dd\sigma_{\rm FC}^{\rm NLO} + \Biggl(a_s~f_{g}\cA\dd\widehat{\sigma}^{(1)}_{gQ\to ZQ}
\\
&+ a_s^2\, f_{g} \cA  \dd\widehat{\sigma}^{(2)}_{gQ\to ZQ(g)}
+a_s^2\,  \Sigma \cA \dd\widehat{\sigma}^{(2)}_{qQ\to ZQ(q)}\Biggr)\cB \delta{f}^{\rm (NLO)}_Q
+(\rm{exch.})\,.
\end{aligned}
\label{eq:tot-residual-pdf2}
\end{equation}

\subsection{A numerical example of bottom-quark PDF residuals \label{sec:ResPDFNumerical}}

\begin{figure}
\centering
\includegraphics[width=0.49\textwidth]{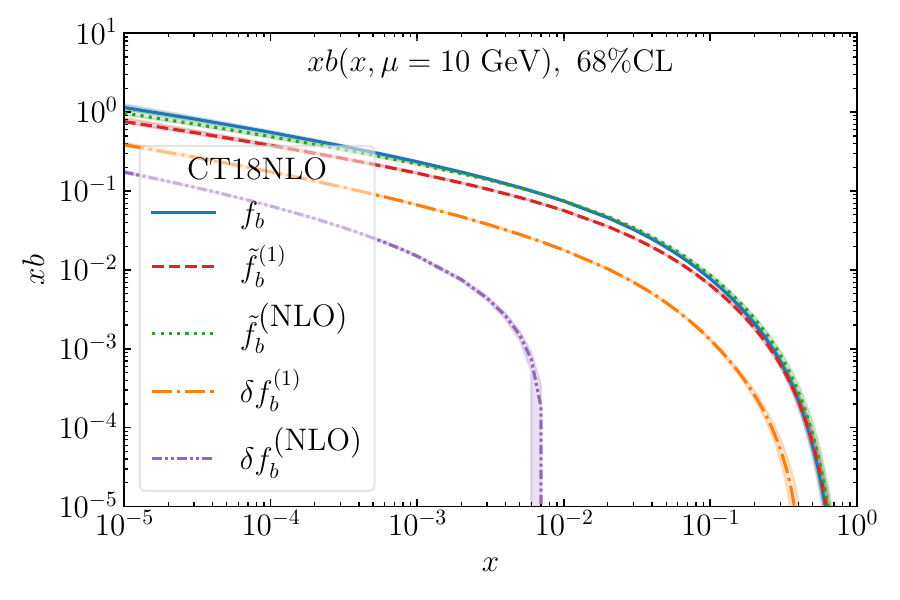}
\includegraphics[width=0.49\textwidth]{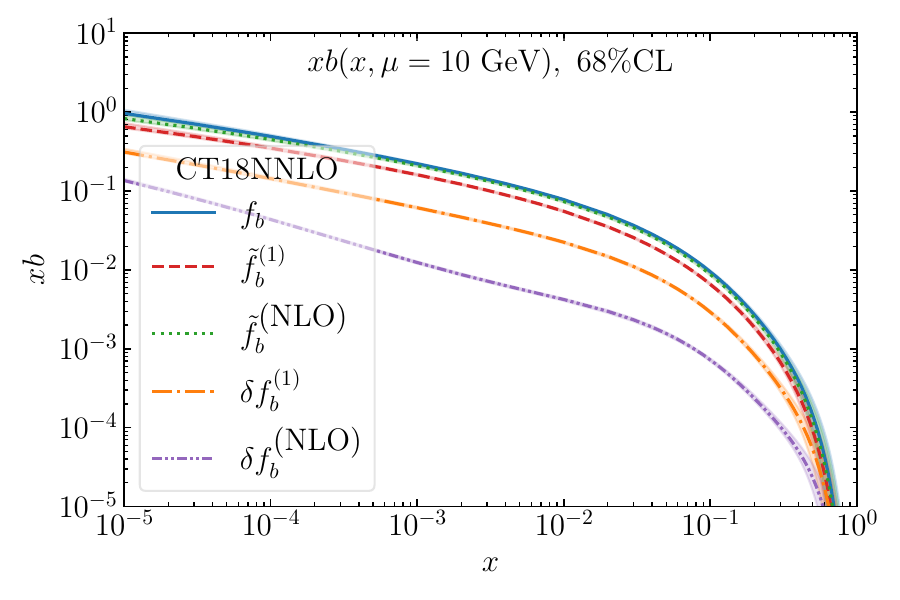}\\
\hspace{20pt}(a) \hspace{3in} (b)\\
\includegraphics[width=0.49\textwidth]{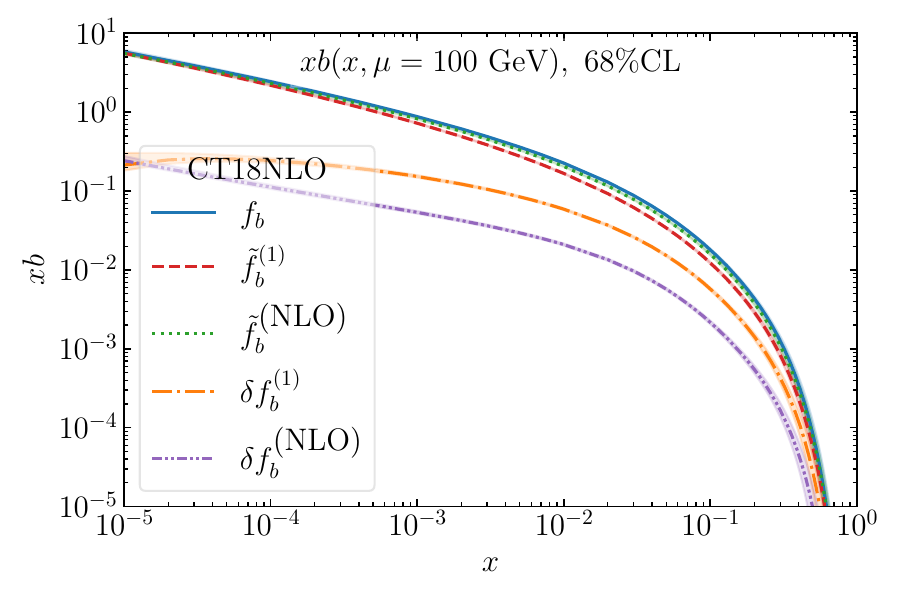}
\includegraphics[width=0.49\textwidth]{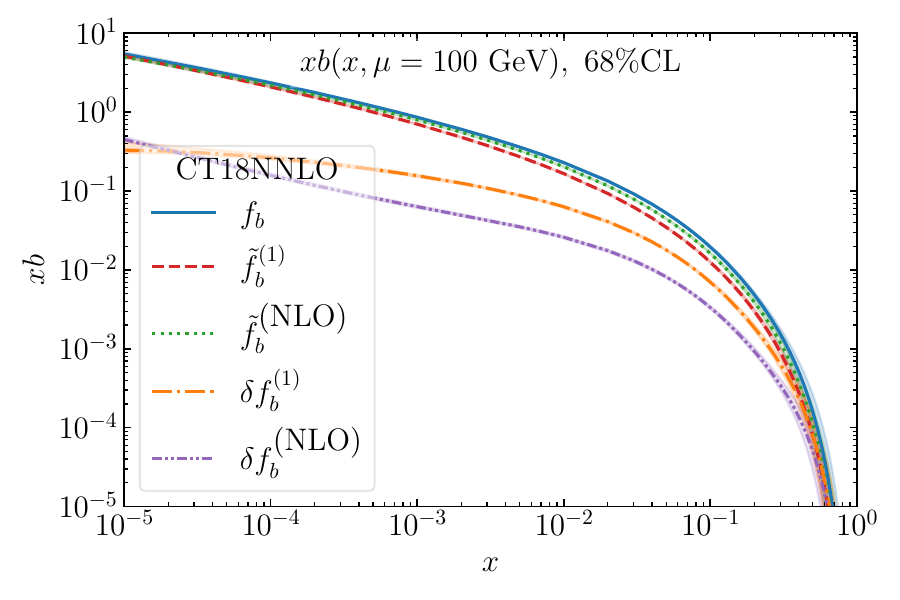}\\
\hspace{20pt}(c) \hspace{3in} (d)\\
\caption{The $b$-quark PDF with PDF subtractions and residuals at the scales $\mu=10$ GeV (upper row) and 100 GeV (lower row). The left (right) column is computed with CT18 NLO (CT18 NNLO) PDFs.
The error bands correspond to the PDF uncertainty evaluated at the 68\% CL.}
\label{fig:SubResPDFsHiQ}
\end{figure}

In Secs~\ref{sec:SubResPDFs} and \ref{sec:Technical}, we constructed the GMVFN cross sections for $Q=b$ by adding the $\dd\sigma_{\rm FE}-\dd\sigma_{\rm sub}$ matching contribution to the flavor-creation contribution $\dd\sigma_{\rm FC}$. The matching contribution is obtained by the convolution of the FE cross section with a special PDF ensemble in which the standard $b$-quark PDF $f_b(x,\mu)$ is replaced by a residual $\delta f_b(x,\mu)$ defined in Eq.~(\ref{eq:res-PDFs-def}) as a difference of $f_b(x,\mu)$ and the PDF subtraction $\tilde f_b(x,\mu)$. The PDFs for other flavors are not modified.
The relevant formulas for the GMVFN cross sections are provided in Eq.~(\ref{eq:tot-residual-pdf}) for the LMO scheme and Eq.~(\ref{eq:tot-residual-pdf2}) in an ACOT-type scheme.
The consistency with the NLO FC cross sections requires the perturbative expansions of the PDF subtractions and residuals to be done up to NLO, as detailed in Sec.~\ref{sec:SubResPDFs}. Within these NLO expansions, either NLO or even NNLO input PDFs can be used within the overall $a_s^3$ accuracy, cf. Sec.~\ref{sec:Technical}.

To streamline the described GMVFN computations, we construct the $b$-quark and antiquark PDF subtractions and residuals at LO ($\tilde{f}_b^{(1)},\delta f_b^{(1)}$) and NLO ($\tilde{f}_b^{\rm (NLO)},\delta f_b^{\rm (NLO)}$) based on the CT18 (N)NLO PDFs~\cite{Hou:2019efy}, with grids in the LHAPDF6 format~\cite{Buckley:2014ana} and publicly available through the HEPForge repository~\cite{sacotmps}. 
The OMEs $A_{Qi}^{(n)}(i=g,q,\bar{q})$ up to $n=2$ in Eqs.~(\ref{eq:sub-PDFs-def}) and (\ref{eq:sub-PDFs-NLO}) are taken from Refs.~\cite{Buza:1995ie,Buza:1996wv}, while the convolution is performed with the HOPPET package~\cite{Salam:2008qg}.
By integrating the FE hard cross sections with the {\it residual} PDF ensemble, one automatically obtains 
the difference $\dd\sigma_{\rm FE}-\dd\sigma_{\rm sub}$ (instead of $\dd\sigma_{\rm FE}$), which then should be added to the $\dd\sigma_{\rm FC}$ cross sections computed with the regular CT18 (N)NLO PDFs, as in Eq.~(\ref{eq:VFN}). 

\subsubsection{Comparisons above the mass threshold}
This approach demonstrates good perturbative convergence at energies of a few tens of GeV typical for $Zb$ production. Figure~\ref{fig:SubResPDFsHiQ} compares the standard $b$-quark PDF, $xf_b$, with the subtractions computed at the LO and NLO, $x\tilde f_b^{(1)}$ and $x\tilde f_b^{\rm (NLO)}$, as well as with the corresponding residuals, $x\delta f_b^{(1)}$ and $x\delta f_b^{\rm (NLO)}$. The upper and lower rows show the comparisons at scales $\mu=$10 (Fig.~\ref{fig:SubResPDFsHiQ} (a) and (b)) and 100 GeV (Fig.~\ref{fig:SubResPDFsHiQ} (c) and (d)), while the left and right columns present these functions for CT18 NLO and CT18 NNLO PDFs, respectively. The error bands indicate the PDF uncertainty evaluated at the 68\% confidence level (CL).

On general grounds, the $b$-quark PDF becomes big enough for the FE contributions to compete with the FC ones when $\mu^2$ is much larger than $m_b^2$, i.e., at $\mu \gtrsim 15$ GeV, and with $x$ being not too large. 
Under these conditions, the $b$-quark PDF is dominated by terms containing powers of 
$L_{m}\equiv\ln\left({\mu^2/m_b^2}\right) \gtrsim 1$ that are summed to all $a_s$ orders by solving the DGLAP equations. 
The upper row in Fig.~\ref{fig:SubResPDFsHiQ} shows that already at $\mu=10$ GeV the perturbative convergence is stable, with the $b$-quark PDF better approximated by the NLO subtraction $\tilde f_b^{\rm (NLO)}$ than by the LO one, $\tilde f_b^{(1)}$. In Fig.~\ref{fig:SubResPDFsHiQ}(a) for CT18 NLO, the NLO subtraction $\tilde f_b^{\rm (NLO)}$ overshoots the $b$-quark PDF at $x>0.007$, resulting in a mildly negative residual $\delta f_b^{\rm (NLO)}$. This does not happen with CT18 NNLO in Fig.~\ref{fig:SubResPDFsHiQ}(b) or at higher $\mu$, with the $b$-quark PDF generally being larger than the subtraction.

\begin{figure}[tb]
\centering
\hspace*{-12pt}
\includegraphics[width=0.34\textwidth]{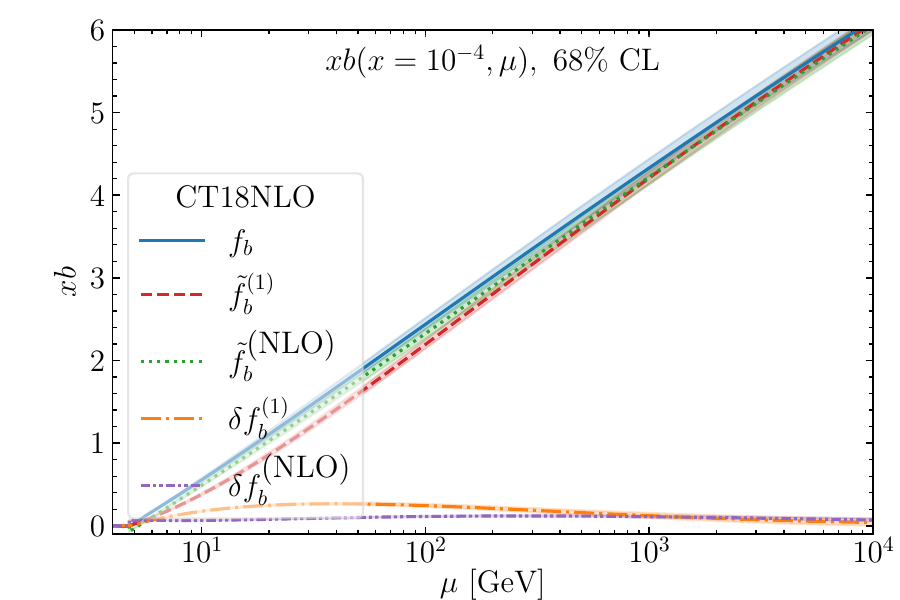}
\hspace*{-7pt}
\includegraphics[width=0.34\textwidth]{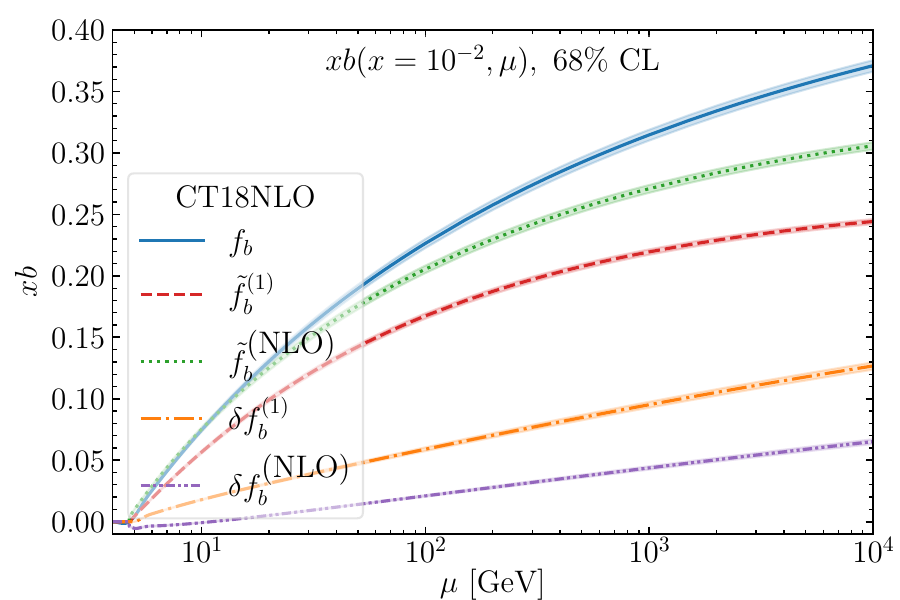}
\hspace*{-7pt}
\includegraphics[width=0.34\textwidth]{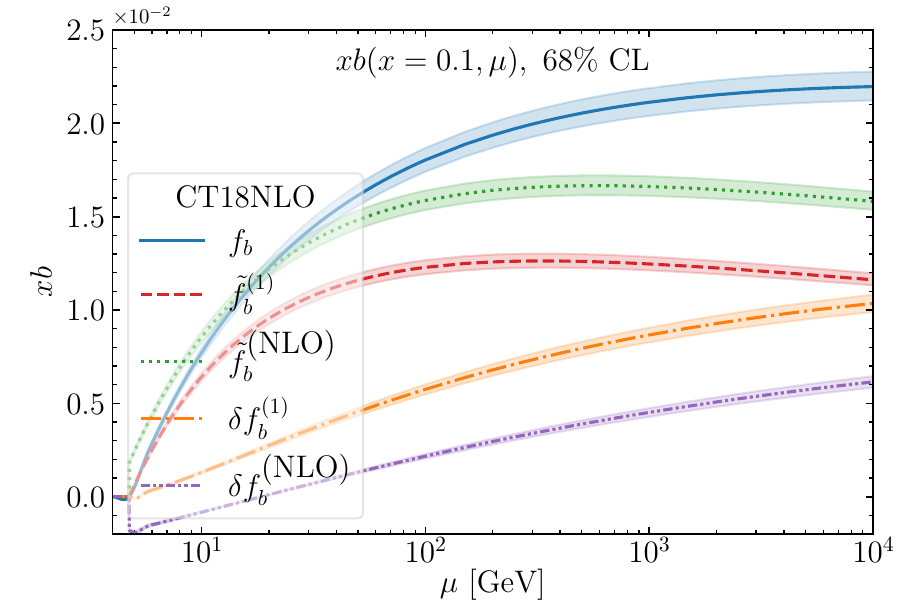}\\
\hspace*{-12pt}
\includegraphics[width=0.34\linewidth]{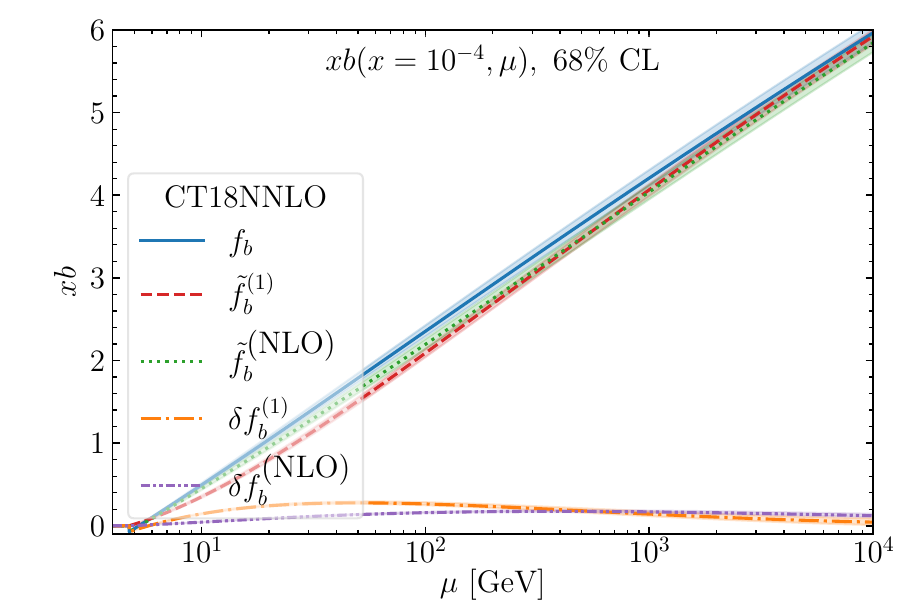}
\hspace*{-7pt}
\includegraphics[width=0.34\linewidth]{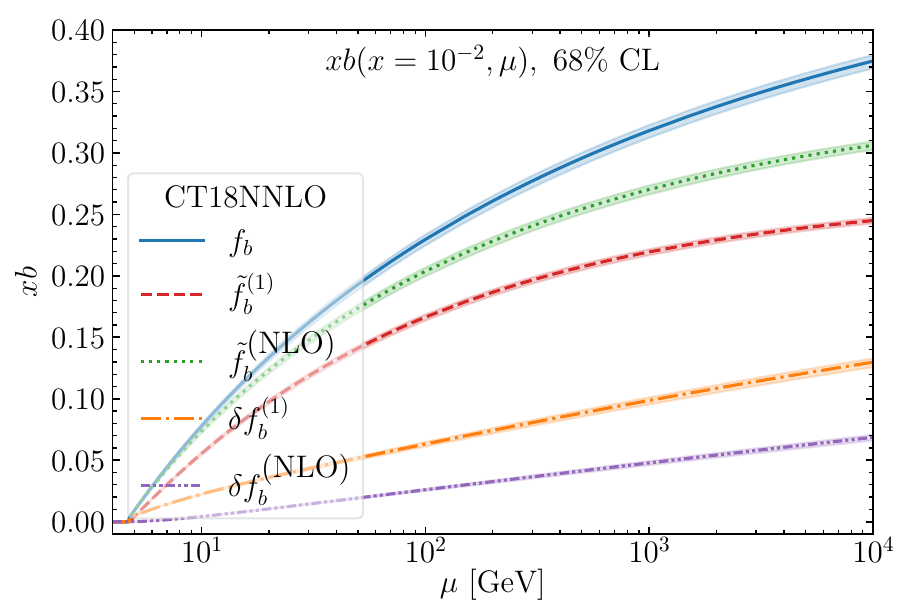}
\hspace*{-7pt}
\includegraphics[width=0.34\linewidth]{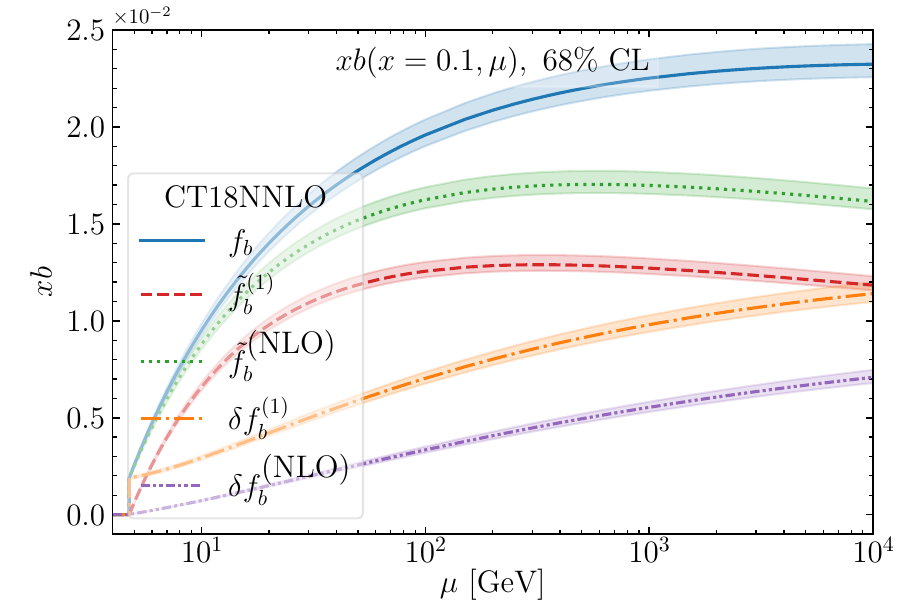}
\caption{The subtraction and residual PDFs in comparison with the $b$-quark PDF at $x=10^{-4}$, $10^{-2}$, and $0.1$ as functions of the scale $\mu$ for CT18NLO (upper row) and CT18NNLO (lower row).}
\label{fig:SubResPDFvsMuHigh}
\end{figure}

\iffalse
\begin{sidewaysfigure}[p]
\centering
\includegraphics[width=0.33\textwidth]{./figures/CT18NLO-Bsub-Bres-nlo-x1em4-05-xbqk-liny.pdf}
\includegraphics[width=0.33\textwidth]{./figures/CT18NLO-Bsub-Bres-nlo-x1em2-05-xbqk-liny.pdf}
\includegraphics[width=0.33\textwidth]{./figures/CT18NLO-Bsub-Bres-nlo-x0p1-05-xbqk-liny.pdf}\\
\includegraphics[width=0.33\linewidth]{./figures/CT18subeps-Bres-nlo-x1em4-05-xbqk-liny.pdf}
\includegraphics[width=0.33\linewidth]{./figures/CT18subeps-Bres-nlo-x1em2-05-xbqk-liny.pdf}
\includegraphics[width=0.33\linewidth]{./figures/CT18subeps-Bres-nlo-x0p1-05-xbqk-liny.pdf}
\caption{The subtraction and residual PDFs in comparison with the heavy-flavor $b$-quark PDFs at $x=10^{-4}$, $10^{-2}$, and $0.1$ as functions of the scale $\mu$ for CT18NLO (upper row) and CT18NNLO (lower row).}
\label{fig:SubResPDFvsMuHigh}
\end{sidewaysfigure}
\fi

Figure~\ref{fig:SubResPDFvsMuHigh} provides another group of comparisons showing the dependence of the $b$-quark, subtraction and residual PDFs on the QCD scale $\mu$ up to 10 TeV for $x=10^{-4}$, $10^{-2}$, and 0.1. At a small momentum fraction, such as $x\sim10^{-4}$, the $b$-quark and subtraction PDFs increase rapidly along with the scale $\mu$. If $x$ decreases more, such as $x<10^{-5}$, we expect the small-$x$ effects to kick in, such as Balitsky--Fadin--Kuraev--Lipatov (BFKL) resummation~\cite{Fadin:1975cb,Kuraev:1976ge,Kuraev:1977fs,Lipatov:1976zz,Balitsky:1978ic} or saturation~\cite{Gribov:1983ivg,Golec-Biernat:1998zce,Morreale:2021pnn}.
The growth with $\mu$ is slower at $x\sim 10^{-2}$, cf. the middle panels of Fig.~\ref{fig:SubResPDFvsMuHigh}.
At even higher $x$, such as $x=0.1$, the monotonic increase of the PDF with $\mu$ eventually gives way to the constant or even decreasing trend, as seen in the rightmost column of Fig.~\ref{fig:SubResPDFvsMuHigh}. At all $x$, the residual PDFs grow with $\mu$, indicating that the $b$-quark PDF includes additional resummed mass logarithms from higher orders that are not present in the respective LO and NLO subtractions.  

\subsubsection{Comparisons near the mass threshold}

\begin{figure}
\centering
\includegraphics[width=0.49\textwidth]{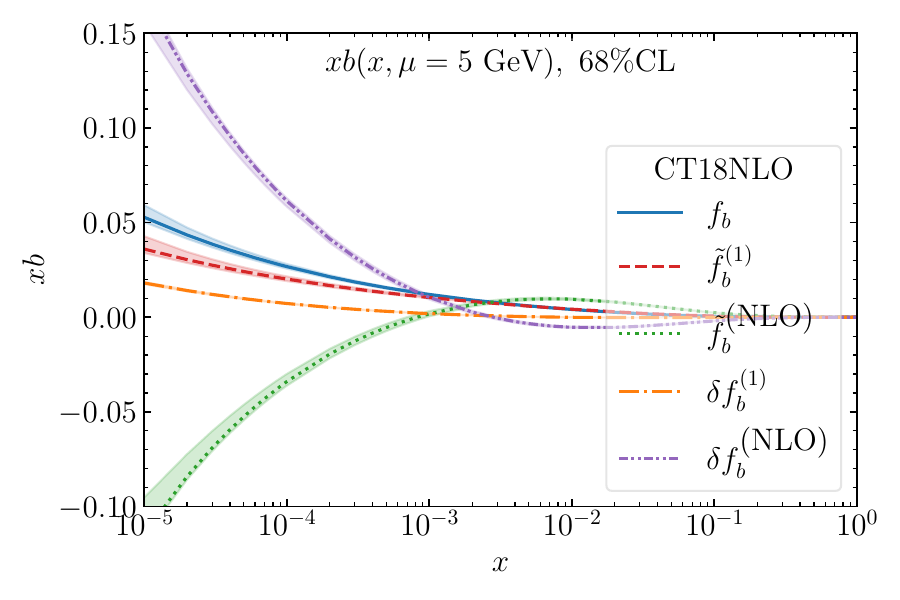}
\includegraphics[width=0.49\textwidth]{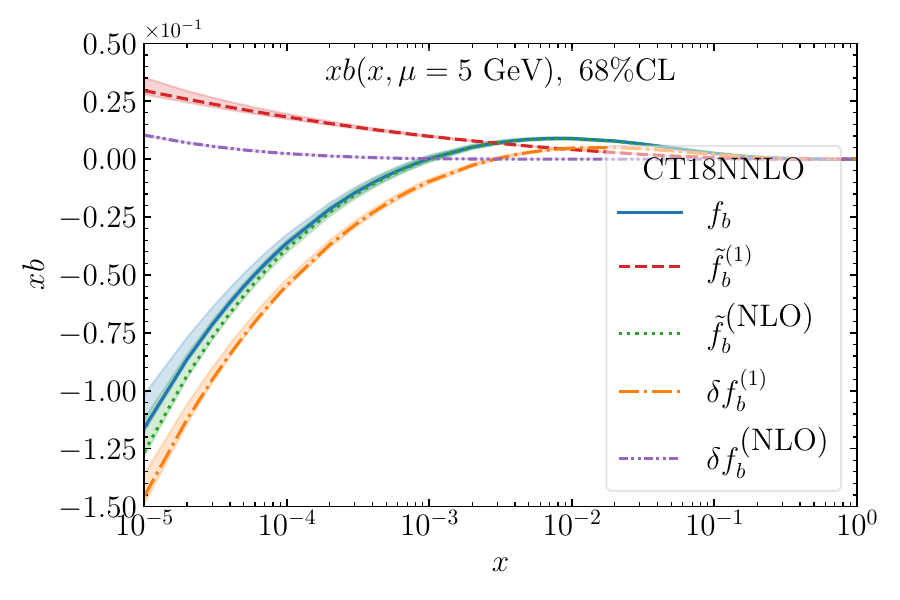}\\
\hspace{20pt}(a) \hspace{3in} (b)
\caption{The $b$-quark PDF with PDF subtractions and residuals at $\mu=5$ GeV for (a) CT18 NLO and (b) CT18 NNLO PDFs.
The error bands correspond to the PDF uncertainty evaluated at the 68\% CL.}
\label{fig:SubResPDFsLowQ}
\end{figure}

We will now review the PDF subtractions and residuals at small scales $\mu\approx m_b$ or $x\to 1$, the regions that are kinematically close to the threshold for production of individual $b$ quarks, but are of less importance for heavier final states, such as $Zb$ production whose partonic production threshold is at approximately 100 GeV. The threshold region is of interest in its own right: here the approximate FE-SUB terms must vanish fast enough to be negligible compared to the physical FC terms. The behavior of FE-SUB is sensitive to the degree of cancellations among the logarithms in the $b$-quark PDFs and subtractions, as well as to
the constant initial contribution to the OMEs at $\mu = m_b$ introduced by switching from $N_f=4$ to $N_f=5$ starting at order $a_s^2$. Furthermore, at $\mu$ above $m_b$, the $b$-quark PDF and subtractions are nonzero at any $x$, while the FE hadronic cross sections vanish below a lower limit on $x_{A,B}$ in Eq.~(\ref{eq:sigma-hadronic}) because of the energy-momentum conservation. Constraints on accessible $x_{A,B}$ must be independently imposed on the FE-SUB {\it hard} cross sections to prevent contributions from small $x$ disallowed in the FE terms. Some heavy-quark schemes, such as S-ACOT-MPS, apply such additional constraints to improve perturbative convergence. But in $Zb$ production with its large mass scales, all these considerations are of less consequence.

In this subsection, we will focus on comparing the threshold behaviors of the $b$-quark PDFs and the subtractions defined at LO or NLO according to Eq.~(\ref{eq:sub-PDFs-def}) and explicitly computed in Eq.~(\ref{sub-with-OMEs2}). Since the order of the subtraction should match the order $a_s^2$ of our FE calculation, our $Zb$ calculation eventually uses the NLO subtraction $\tilde{f}_b^{\rm(NLO)}$, which, however, can be evaluated using NLO PDFs at the lowest mandatory order or NNLO PDFs in the other prescriptions like ACOT. A subtlety to notice is that the $A^{(2)}_{Qj}$ OME in this subtraction includes a constant term due to $N_f=4\to 5$ switching, but the common PDFs like CT18 include this constant only starting from NNLO. Specifically, the OMEs $A^{(2)}_{Qj}$ in $\tilde{f}_b^{\rm(NLO)}$ contain single and double powers of $L_{m}\equiv\ln\left({\mu^2/m_b^2}\right)$, denoted by $a_{Qj}^{(2,1)}L_{m}$ and $a_{Qj}^{(2,2)}L^2_{m}$, as well as a constant matching term $a_{Qj}^{(2,0)}$ that is independent of $\mu$. While the logarithmic contributions vanish at $\mu^2\to m_b^2$, the matching term does not: it drives the nonzero asymptotics for $\tilde{f}_b^{\rm(NLO)}$ at $\mu^2\to m_b^2$. No such term is present at the lower order, i.e., $\tilde{f}_b^{(1)}$ tends to zero. The $a_{Qj}^{(2,0)}$ term thus determines which subtraction order better matches the $b$-quark PDF computed with either CT18 NLO or CT18 NNLO when approaching the threshold.

As an illustration, Fig.~\ref{fig:SubResPDFsLowQ} plots the $b$-quark PDFs, subtractions, and residuals at $\mu=5~\GeV$, chosen slightly above the pole bottom-quark mass $m_b=4.75~\GeV$ adopted in CT18. We see that these PDFs are particularly stable (and small) at $x\gtrsim 10^{-3}$ for both CT18 NLO in the left panel and CT18 NNLO in the right one. Although the LO subtraction $\tilde{f}^{(1)}_b$ (red) and NLO one $\tilde{f}^{\rm (NLO)}_b$ (green) have different shapes, they are both close to the $b$-quark PDFs, resulting in small residuals $\delta{f}^{(1)}_b$ (orange) and $\delta{f}^{\rm (NLO)}_b$ (green) at the respective orders at $x\gtrsim 10^{-3}$.

More variability is apparent at $x < 10^{-3}$, where $\tilde{f}^{(1)}_b$ (red) and $\tilde{f}^{\rm (NLO)}_b$ (green) are quite distinct in both panels of Fig.~\ref{fig:SubResPDFsLowQ}. In the left panel for CT18 NLO, the $b$-quark PDF (blue) is closer to $\tilde{f}^{(1)}_b$, while in the right panel for CT18 NNLO, it is closer to $\tilde{f}^{\rm (NLO)}_b$. Consequently, the smaller of the two residuals is $\delta{f}^{(1)}_b$ in Fig.~\ref{fig:SubResPDFsLowQ}(a) and $\delta{f}^{\rm (NLO)}_b$ in Fig.~\ref{fig:SubResPDFsLowQ}(b). This difference arises from the matching term $a_{Qj}^{(2,0)}$ included in $\tilde{f}^{\rm (NLO)}_b$ and CT18 NNLO and not in $\tilde{f}^{(1)}_b$ and CT18 NLO.

\begin{figure}[tb]
\centering
\hspace*{-12pt}
\includegraphics[width=0.49\textwidth]{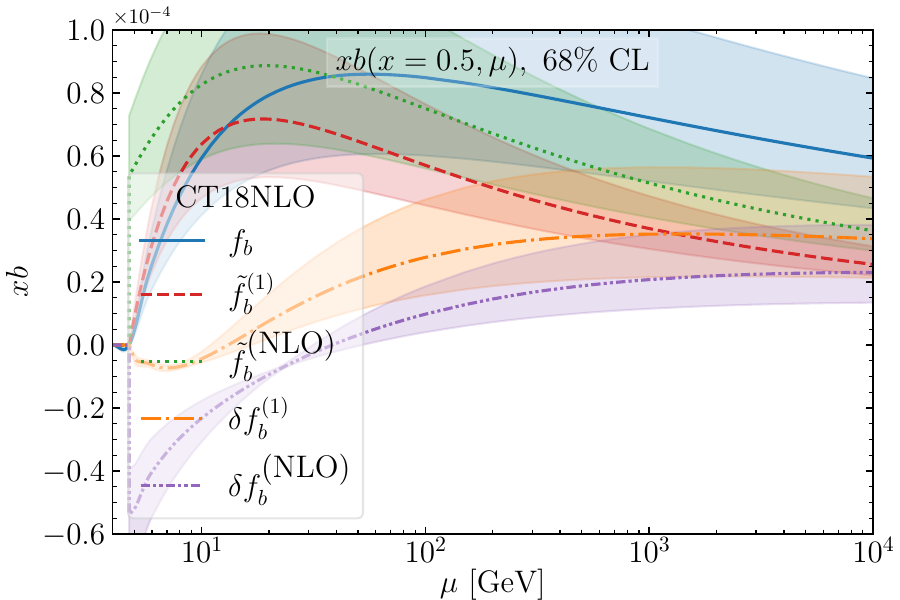}
\includegraphics[width=0.49\linewidth]{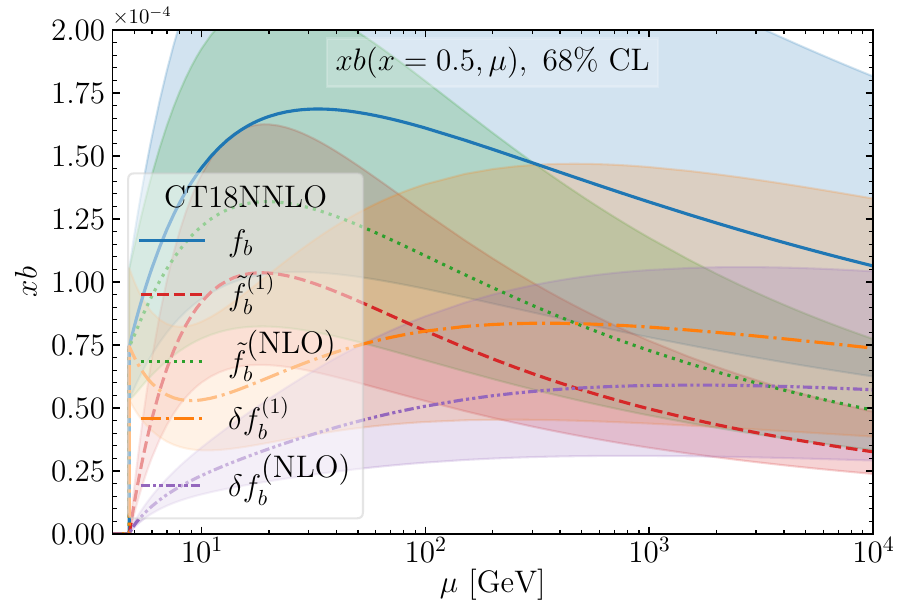}\\
\hspace{20pt}(a) \hspace{3in} (b)
\caption{Similar to Fig.~\ref{fig:SubResPDFvsMuHigh}, for $x=0.5$ for (a) CT18 NLO and (b) CT18 NNLO.}
\label{fig:SubResPDFvsMuLow}
\end{figure}

At higher scales, the logarithmic terms introduced by DGLAP evolution drive the growth of the $b$-quark PDF and subtractions, as was already demonstrated in Figs.~\ref{fig:SubResPDFsHiQ} and \ref{fig:SubResPDFvsMuHigh}. 
In the latter figure, one can notice the discontinuities at $\mu=m_b$ arising from the nonlogarithmic term $a_{Qj}^{(2,0)}$ associated with the $N_f=4\to 5$ switching at representative values of $x$. The impact of the discontinuity is limited to the smallest $\mu$. In fact, the discontinuity is most visible for $x=0.1$ and barely noticeable at $x=10^{-4}$ and $10^{-2}$.
At larger $\mu$, it is rapidly overtaken by the logarithmic terms in the PDFs. 
The discontinuity is reduced with CT18 NNLO PDFs. Either way, the differences between the NLO residuals of CT18 NLO and NNLO PDFs are weak and limited to $\mu$ of order a few tens GeV,  far below the typical scales of $Zb$ production. The nonlogarithmic term $a_{Qj}^{(2,0)}$ in $A^{(2)}_{Qj}$ thus quickly becomes negligible compared to the logarithmic contributions, except at very large $x$ where the logarithmic growth is suppressed. This point is illustrated by Fig.~\ref{fig:SubResPDFvsMuLow}, showing the $\mu$ dependence of the PDFs at $x=0.5$, together with the bands indicating the PDF uncertainties. While the matching discontinuities are visible for the plotted PDFs in both panels, the discontinuity of the NLO residual is subdued when using CT18 NNLO in the right panel.

\section{Sample results for $Z+b$ jet production at the LHC} 
\label{sec:results}

\begin{table}[p]
\centering
\begin{tabular}{c|c|c|c|c||c|c|c|c}
\hline
 $\sigma_{\rm FC}^{\rm{LO,NLO}}$(pb) & \multicolumn{4}{c||}{fixed $\mu$} & \multicolumn{4}{c}{dynamic $\mu$} \\ \hline
Process & LO $b$& LO $b\bar b$& NLO $b$& NLO $b\bar b$ & LO $b$ & LO $b\bar b$ & NLO $b$ & NLO $b\bar b$\\ \hline
$gg$ & $205^{+49}_{-38}$ & $1.6^{+0.3}_{-0.3}$& $394^{+26}_{-34}$ & $2.6^{+0}_{-0.1}$ & $195^{+46}_{-35}$ & $1.5^{+0.3}_{-0.3}$ & $385^{+29}_{-34}$ & $2.6^{+0}_{-0.1}$ \\
$q\bar q$ & $24^{+4}_{-3}$& $5.1^{+0.8}_{-0.7}$& $36.7^{+0}_{-0.8}$ & $7.9^{+0}_{-0.2}$ & $23^{+4}_{-3}$& $4.8^{+0.7}_{-0.6}$ & $37^{+0}_{-1}$& $7.9^{+0}_{-0.3}$\\
$qg+\bar q g$& -& -& $12^{+38}_{-20}$ & $9^{+5}_{-3}$ & - & -& $5^{+32}_{-17}$& $7^{+4}_{-2}$\\ \hline
$pp \to Zb \bar b$& $229^{+53}_{-41}$& $6.7^{+1.1}_{-0.9}$ & $443^{+63}_{-55}$ & $19^{+5}_{-3}$ & $218^{+49}_{-38}$ & $6.4^{+1.0}_{-0.9}$ & $427^{+61}_{-52}$ & $18^{+4}_{-3}$\\ \hline
\end{tabular}
\caption{Total cross sections for FC contributions at LO and NLO QCD for the production of $Z+$ at least one $b$ jet via flavor creation processes with $m_b=4.75$ GeV at the 13 TeV LHC. The reported errors are purely from scale
variation. Separately listed are the contributions from the partonic subprocesses for $gg, q\bar q$, and $qg+\bar q g$ ($q=u,d,s,c$) initiated $Zb \bar b$ production and their sum as described in Eq.~(\ref{eq:tot-FC}). 
Since $b $jets can originate from one $b$ (or $\bar b$) quark (labeled as LO $b$, NLO $b$) or from the combination of a $b$ and $\bar b$ (labeled as LO $b\bar b$, NLO $b\bar b$), the corresponding cross sections are provided separately. See text for a detailed discussion.}
\label{tab:FC-results}
\end{table}

\begin{table}[p]
\begin{center}
\begingroup
\footnotesize
\begin{tabular}{c|c|c|c||c|c|c||c|c|c||c|c|c}
\hline
$\sigma_{\rm FE,sub}$ (pb) & \multicolumn{3}{c||}{ACOT, fixed $\mu$} & \multicolumn{3}{c||}{ACOT, dynamic $\mu$} 
& \multicolumn{3}{c||}{S-ACOT, fixed $\mu$} & \multicolumn{3}{c}{S-ACOT, dynamic $\mu$}
\\ \hline
Process & LO & NLO & sub & LO & NLO & sub & LO & NLO & sub & LO & NLO & sub \\ \hline
$bg$ & $388^{+15}_{-31}$ & $557_{-99}^{+75}$ & $476_{-69}^{+48}$ & $391^{+12}_{-27}$ & $545^{+91}_{-79}$  &  $486^{+44}_{-64}$ & $391^{+15}_{-41}$ & $535_{-107}^{+73}$ & $460_{-68}^{+48}$ & $393^{+13}_{-28}$ & $516^{+91}_{-86}$ & $469^{+45}_{-63}$  \\
$bq+b\bar q$ &- & $38_{-20}^{+22}$  & $28_{-15}^{+17}$ & - & $33^{+22}_{-20}$ & $24^{+17}_{-15}$ 
&- & $38_{-21}^{+22}$ & $28_{-15}^{+17}$ & - & $33^{+22}_{-20}$ & $24^{+17}_{-15}$ \\  \hline
sum & $388_{-31}^{+15}$ & $595_{-77}^{+55}$ & $504_{-52}^{+33}$  & $391^{+12}_{-27}$ & $578^{+72}_{-57}$&  $511^{+29}_{-48}$
& $391^{+15}_{-41}$ & $573_{-85}^{+53}$ & $488_{-51}^{+33}$ & $393^{+13}_{-28}$ & $549_{-64}^{+72}$ & $493_{-46}^{+30}$ \\ \hline
\end{tabular}
\endgroup
\end{center}
\caption{Total cross sections for FE and NLO subtraction terms for production of $Z+$ at least one $b-$jet at LO and NLO QCD. The reported errors are purely from scale variation. Separately listed are the contributions from the partonic subprocesses for $bg$ and $bq+b\bar q$ $(q=u,d,s,c)$ initiated $Zb$ production, and their sum as described in Eq.~(\ref{eq:tot-FE}). See text for a detailed discussion. }
\label{tab:FE-results}
\end{table}

\begin{table}[p]
\centering
\begin{tabular}{c|c|c||c|c}
\hline
$\sigma_{\rm GMVFN}^{\rm NLO}$ (pb)  & ACOT & S-ACOT & ACOT & S-ACOT \\ 
 \hline
 & \multicolumn{2}{c||}{$b$ case} & \multicolumn{2}{c}{$b+b \bar b$ case}  \\ \hline
Fixed $\mu$ & $535^{+38}_{-32}$ & $528^{+29}_{-35}$ & $554^{+43}_{-36}$  & $547^{+33}_{-38}$ \\
Dynamic $\mu$ & $494^{+52}_{-9}$ & $483^{+43}_{-10}$ & $512^{+56}_{-12}$ & $501^{+47}_{-13}$ \\ \hline
\end{tabular}
\caption{Full GMVFN cross sections (FC+FE-SUB) for production of $Z+$ at least one $b-$jet obtained from the NLO cross sections in Tables~\ref{tab:FC-results} and \ref{tab:FE-results}. The reported errors are purely from scale variation. 
}
\label{tab:GMVFN-results}
\end{table}

In this section, we present results for the production of a $Z$ boson  with at least one $b$ jet at the LHC with $\sqrt{s}=13$~TeV, as obtained from Eqs.~(\ref{eq:tot-FC}), (\ref{eq:tot-FE}), and (\ref{eq:tot-subtraction-pdf}). This corresponds to the LMO implementations including NLO QCD corrections in the ACOT and S-ACOT schemes. As previously explained, while in the FC case $b$ quarks are always treated as massive, for the $\dd\sigma_{\rm FE}-\dd \sigma_{\rm sub}$ terms we consider both the expressions with a massive (ACOT LMO) and a massless (S-ACOT LMO) $b$ quark. 

As input parameters, we choose
\begin{equation}
M_Z=91.1876~\GeV,~M_W=80.379~\GeV, ~ 
G_F=1.1663787\times 10^{-5}~\GeV^{-2}.
\end{equation} 
We work with the CT18NLO PDF set~\cite{Hou:2019efy} with NLO $\alpha_s(M_Z)=0.118$, and use the set of subtraction PDFs defined in Sec.~\ref{sec:SubResPDFs}. In the massive case, we set the $b$-quark pole mass to $m_b=4.75~\GeV$, consistently with the CT18 choice, and adopt consistent fully massive kinematics (see~\cite{Figueroa:2018chn}
for details).

We have considered both the case of fixed and dynamic renormalization and factorization scales. In both cases we indicatively estimate the uncertainty due to missing higher orders by varying $\mu=\mu_R=\mu_F$ by a factor of two around the central value that is picked to be $\mu_0=M_Z$ in the fixed-scale case and $\mu_0=M_{T,Z}$ in the dynamic-scale case, where the $Z$-boson transverse mass ($M_{T,Z}$) is defined in terms of its transverse momentum $p_{T,Z}$ as $M_{T,Z}=\sqrt{M_Z^2+(p_{T,Z})^2}$.

We reconstruct $b$ jets using a $k_T$ algorithm with $R=0.4$ (where $R=\sqrt{\Delta\phi^2+\Delta\eta^2}$ is the radius of the jet in the azimuthal angle-pseudorapidity plane) and, following choices commonly made by the LHC experiments, we identify or \textit{tag} $b$ jets by imposing the following cuts on transverse momentum and pseudorapidity:
\begin{equation}\label{eq:cut}
p_{T,b}>25~\GeV,~|\eta_b|<2.5.
\end{equation}
In our parton-level calculation, $b$ jets can be formed from the recombination of at most three partons and can contain either one $b$ or $\bar{b}$ quark, or a pair of $b\bar{b}$ quarks.
The combined $b\bar{b}$ jet can be declared either as a $b$ jet~\cite{Campbell:2008hh} (an experimental-driven definition),
or as an unflavored jet as proposed in Refs.~\cite{Banfi:2006hf,Czakon:2022wam,Caola:2023wpj} (a theoretical infrared-safe definition adopted in the $W+c$~\cite{Czakon:2020coa,Gehrmann-DeRidder:2023gdl} and $Z+b$~\cite{Gauld:2020deh,Mazzitelli:2024ura} NNLO calculations). In order to facilitate the use of the results presented in this paper, we break down our results for $Z$-boson production with at least one $b$ jet in all their components, and present LO and NLO QCD total cross sections for the case of  FC ($Zb\bar{b}$) in Table~\ref{tab:FC-results} and for the case of FE ($Zb$) with a massive and massless $b$ quark in Table~\ref{tab:FE-results}, including the contribution of the individual partonic channels. Furthermore, in the case of FC we provide results for both the case where the tagged $b$ jet contain one $b$ or $\bar{b}$ quark, as well as for the case where it contains a $b\bar{b}$ pair. We keep all
events that have at least one $b$ jet. For events with 2 $b$ jets, the
differential distributions for $b$-jet observables show the hardest $b$ jet. In both NLO and LO results we use NLO PDFs and the two-loop running of $\alpha_s$ evolved with $N_f=5$. The results for the GMVFN total cross sections at NLO in QCD for the production of $Z+$ at least one $b$ jet in the ACOT LMO and S-ACOT LMO implementation at the 13 TeV LHC are reported in Table \ref{tab:GMVFN-results}. These are obtained according to Eqs.~(\ref{eq:VFN})-(\ref{eq:tot-subtraction-pdf}) and with the adjustments described in Sec.~\ref{sec:Technical} point a). It is interesting to compare the ACOT result in Table \ref{tab:GMVFN-results} with the result obtained in the massive 5-flavor scheme (m5FS) of Ref.~\cite{Figueroa:2018chn}. In Ref.~\cite{Figueroa:2018chn} FC and subtraction cross sections are calculated at LO. Using $\sigma_{\rm FC}^{\rm LO}$ as provided in Table~\ref{tab:FC-results} and $\sigma_{\rm sub}^{\rm LO}=285^{+5}_{-18}$ ($286^{+3}_{-15}$) pb for the fixed(dynamic) scale choice we find for the $b+b\bar b$ case $546^{+9}_{-5}$ pb ($517^{+30}_{-0}$ pb). 

\begin{figure}
\includegraphics[width=0.49\textwidth]{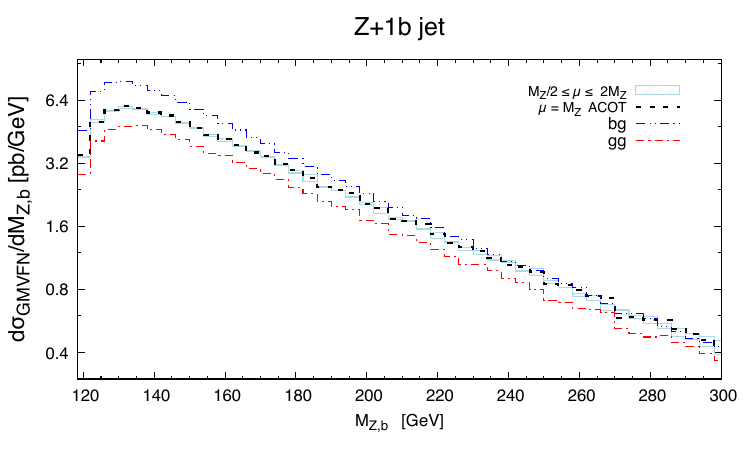}
\includegraphics[width=0.49\textwidth]{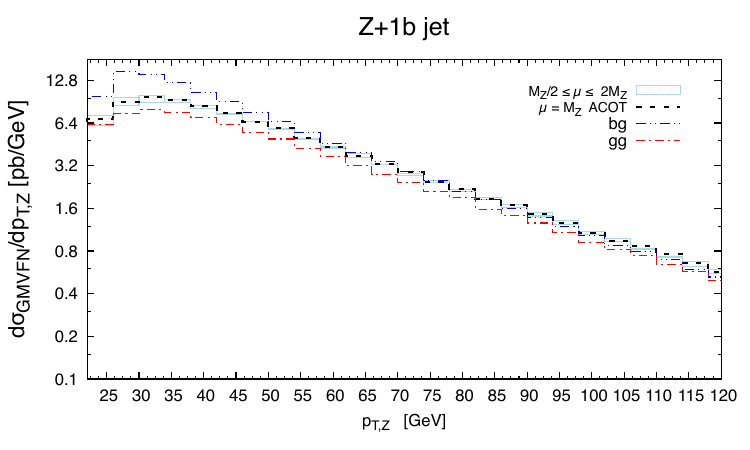}
\includegraphics[width=0.49\textwidth]{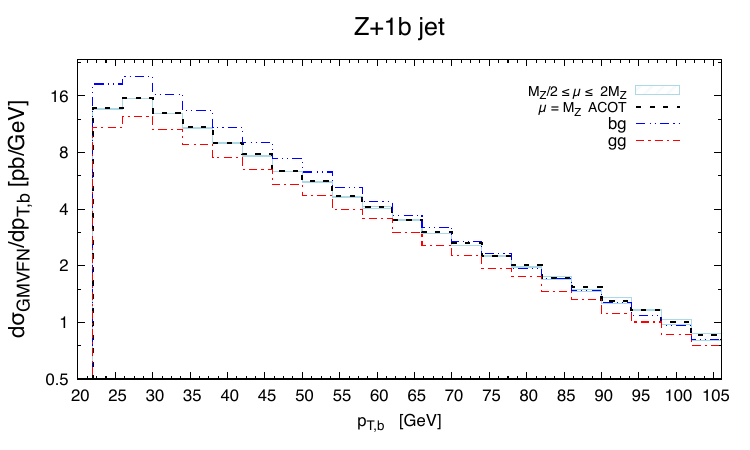}
\includegraphics[width=0.49\textwidth]{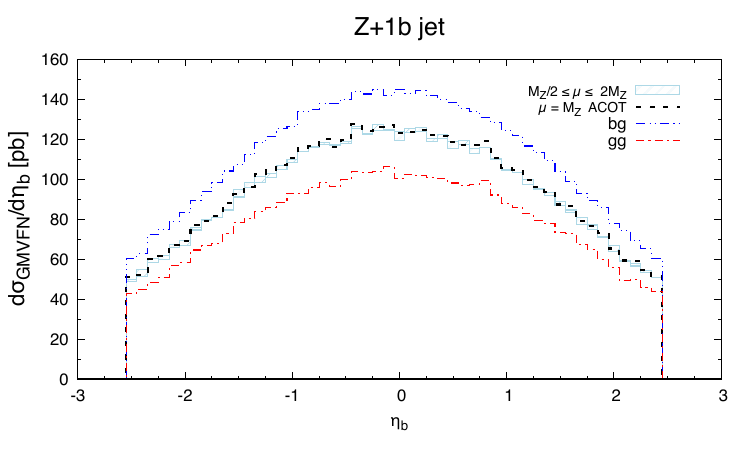}
\caption{Differential distributions with scale dependence $M_Z/2<\mu<2M_Z$ at order ${\cal O}(\alpha_s^3)$ for the ACOT GMVFN scheme in the $gg$ channel as reflected in Fig.~\ref{fig:VFNgg}. The CT18NLO PDFs are used.}
\label{gg:NLO}
\end{figure}

\begin{figure}
\includegraphics[width=0.49\textwidth]{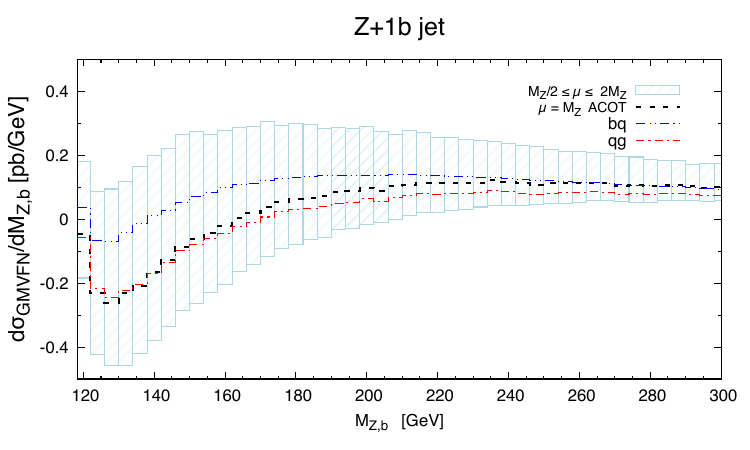}
\includegraphics[width=0.49\textwidth]{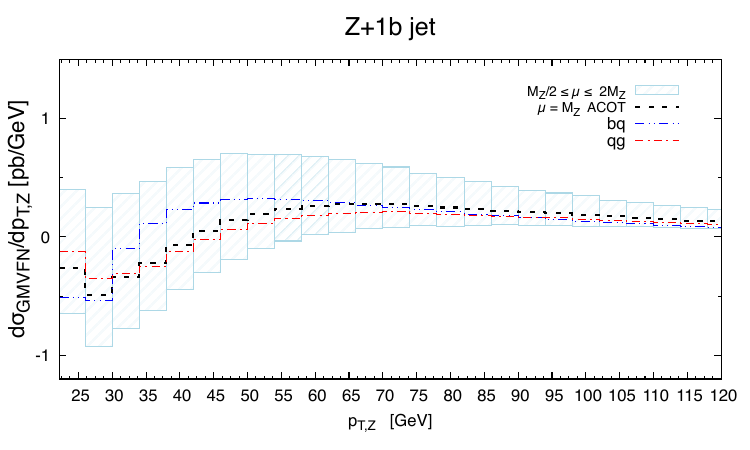}
\includegraphics[width=0.49\textwidth]{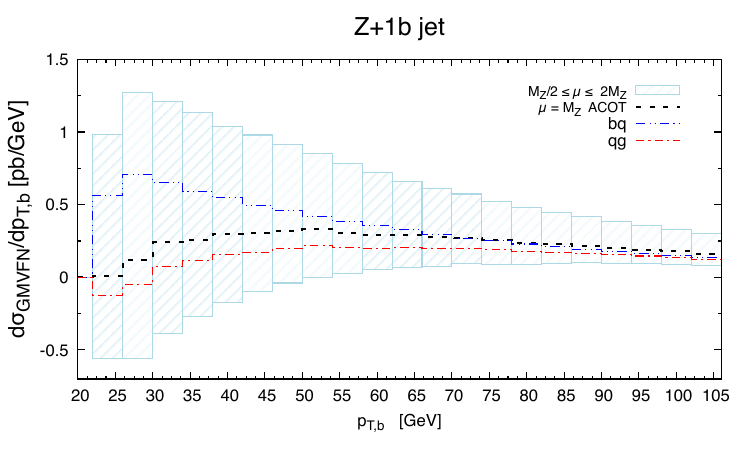}
\includegraphics[width=0.49\textwidth]{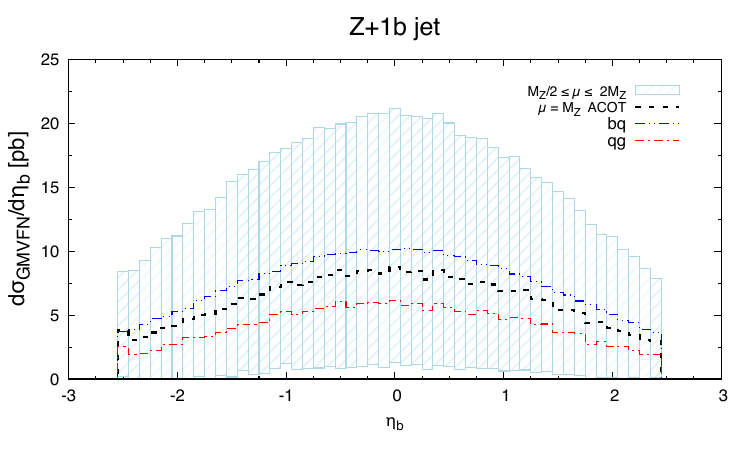}
\caption{Same as in Fig.~\ref{gg:NLO} but in the $qg$ channel as reflected in Fig.~\ref{fig:VFNqg}.}
\label{qg:NLO}
\end{figure}

\begin{figure}
\includegraphics[width=1.0\textwidth]{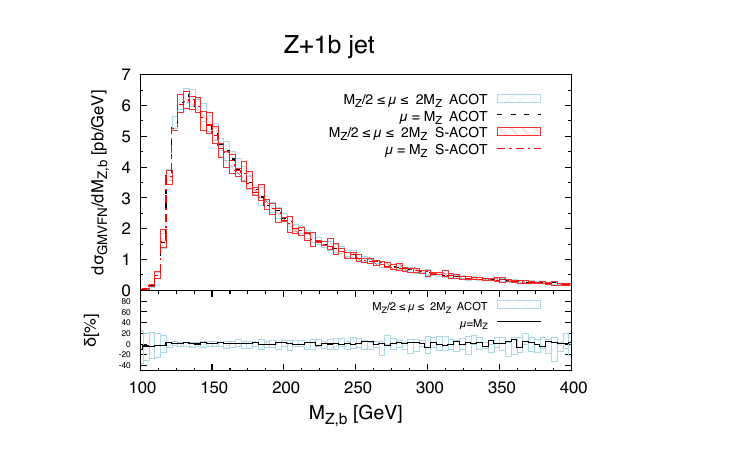}
\includegraphics[width=1.0\textwidth]{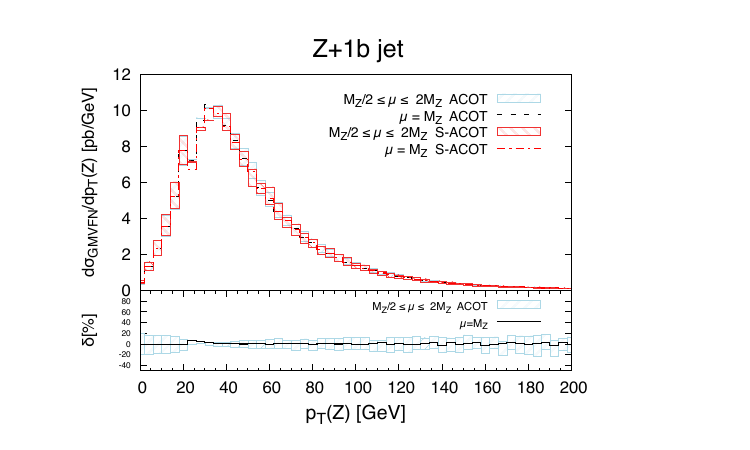}
\caption{$M_{Z,b}$ and $p_{T,Z}$ differential distributions with scale dependence $M_Z/2<\mu<2M_Z$ at order ${\cal O}(\alpha_s^3)$ for the ACOT and S-ACOT GMVFN schemes. The CT18NLO PDFs are used.}
\label{acot-vs-sacot:MZb-ptZ-NLO}
\end{figure}

\begin{figure}
\includegraphics[width=1.0\textwidth]{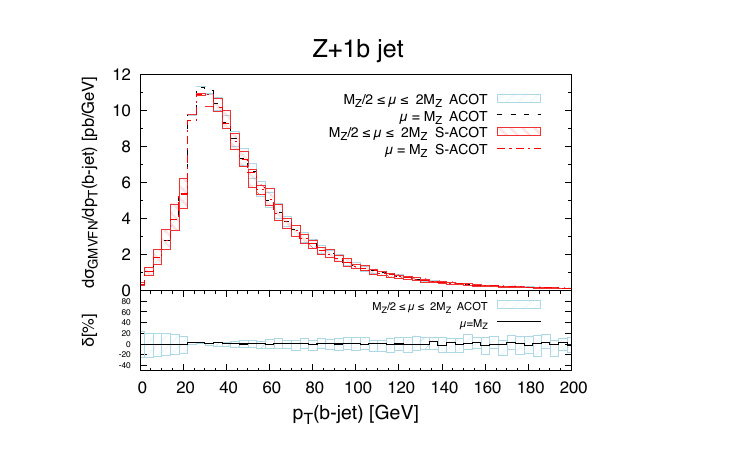}
\includegraphics[width=1.0\textwidth]{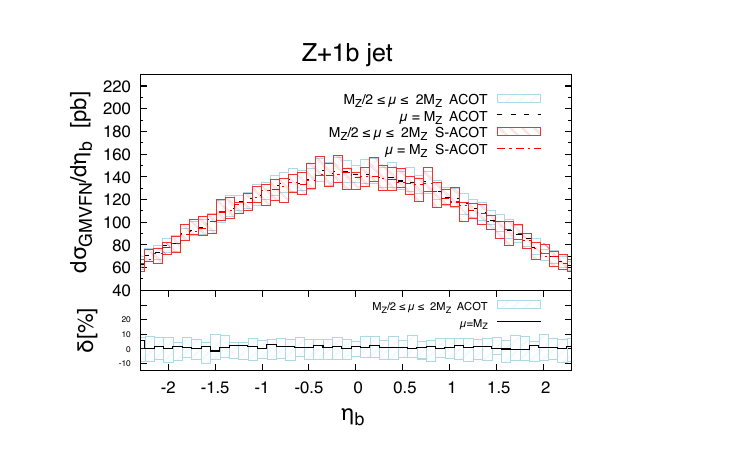}
\caption{$p_{b}$ and $\eta_{b}$ differential distributions with scale dependence $M_Z/2<\mu<2M_Z$ at order ${\cal O}(\alpha_s^3)$ for the ACOT and S-ACOT GMVFN schemes. The CT18NLO PDFs are used.}
\label{acot-vs-sacot:ptb-etab-NLO}
\end{figure}

Theory prediction results for the invariant mass of the $Z+b$ system $M_{Z,b}$, the transverse momentum of the $Z$-boson $p_{T,Z}$, the transverse momentum of the $b$-quark $p_{T,b}$, and the $b$-quark pseudorapidity $\eta_{b}$ distributions, are shown in Figs.~\ref{gg:NLO}-\ref{acot-vs-sacot:ptb-etab-NLO}. In Figs.~\ref{gg:NLO} and~\ref{qg:NLO}, we illustrate the interplay between FC, FE and subtraction terms, separately for the $gg$-FC channel (Fig.~\ref{gg:NLO}) as reflected in Fig.~\ref{fig:VFNgg} in Sec.~\ref{sec:ZQ-hard-scattering}, and for the $qg$-FC channel (Fig.~\ref{qg:NLO}) as reflected in Fig.~\ref{fig:VFNqg} in the same section. 
The ACOT-$gg$($qg$) central prediction is represented by a black dashed line, while its scale dependence $M_Z/2\leq \mu\leq 2 M_Z$ is represented by a light blue band. The FC and FE terms are represented by a red dot-dashed and a blue dot-dot-dashed line, respectively. In both the $gg$- and $qg$-FC channels, we observe that the ACOT central prediction approaches the FE one at large scales and the FC prediction near the threshold, as expected. 
The convergence appears to be faster in the $qg$ channel, where the FE cross section is largely canceled by the subtraction terms at $M_{Z,b}\approx 120$ GeV and at $p_T\approx$ 25 GeV, in contrast to the $gg$ channel, where the matching of terms seems to be more sensitive to phase-space integration and applied cuts. In the case of the $\eta_b$ distribution, where the $p_T$ and other energy scales are integrated out, the central ACOT-like prediction lies in between the FC and FE contributions and within the scale uncertainty in both the $gg$ and $qg$ channels.  
The scale dependence in the $qg$ channel is much larger as compared to that of the $gg$ channel due to the different weights of terms in the logarithmic structure of the theory prediction for these two channels. In fact, the $qg\rightarrow Z b \bar b + q$ appears for the first time at NLO ${\cal O}(\alpha_s^3)$ in the perturbative expansion of the $Z b \bar b$ cross section. The $gg$-channel contributes for most of the cross section (e.g., 85\%-90\%), while the scale dependence is almost totally induced by that in the $qg$ channel (see Figs.~\ref{acot-vs-sacot:MZb-ptZ-NLO} and \ref{acot-vs-sacot:ptb-etab-NLO}).       

In Figs.~\ref{acot-vs-sacot:MZb-ptZ-NLO} and \ref{acot-vs-sacot:ptb-etab-NLO}, we compare the ACOT and S-ACOT theory predictions within their corresponding scale uncertainty bands (light blue for ACOT and light red for S-ACOT). In the lower inset of each distribution, $\delta[\%]$ represents the percent difference between S-ACOT and ACOT relative to ACOT, within the scale uncertainty of the ACOT prediction. Differences are in general around 2-3\% and smaller for all distributions, but they can be larger at higher values of the hard scale of the process. However, all differences are well within the ACOT scale dependence.

\section{Outlook and conclusions}
\label{sec:outlook}

In this work, we presented an application of ACOT-like GMVFN schemes to proton-proton collisions in which at least one heavy quark is produced. The necessary theory framework to generalize these schemes to the case of hadron-hadron reactions is developed based on QCD factorization. This framework is applied to $Z$-boson production in association with a least one $b$-quark jet at ${\cal O}(\alpha_s^3)$ (NLO) in QCD as an illustrative case. We presented results for the total cross section at the LHC with a collision energy of $\sqrt{s}=13$ TeV, for which we explored fixed and dynamical scale dependence, as well as differences between the ACOT LMO and S-ACOT LMO schemes. In addition, we studied several differential distributions and their scale dependence, such as the invariant mass of the $Z+b$ system, the transverse momentum $p_{T,Z}$ of the $Z$ boson, the transverse momentum $p_{T,b}$ and the pseudorapidity $\eta_b$ of the $b$ quark. 

The practical implementation of the theory calculation for the $Z+b$ cross section is facilitated by introducing the concepts of subtraction and residual HQ PDFs. The former consists of convolutions between PDFs and universal operator matrix elements representing the transition from a massless parton to a heavy quark, including mass dependence and expanded to the same $\alpha_s$ order as the flavor-creation cross section. The latter is the difference of the $b$-quark PDF and subtraction PDF. Tabulated grids for CT18 NLO and CT18 NNLO PDF ensembles, in which the $b$-quark PDFs are replaced by either the residual or subtraction $b$-quark PDFs, are provided in the LHAPDF6 format~\cite{Buckley:2014ana} and distributed through a repository at HEPForge~\cite{sacotmps}. These grids are process independent and can be applied to construct ACOT-like GMVFN theory predictions for other processes.    

Our discussion focused on $b$-quark mass effects in the $Z+b$ process in proton-proton reactions, while the $c$ quark was treated as massless together with the other light quarks for simplicity. However, ACOT-like factorization schemes are very general and can simultaneously account for $c$- and $b$-quark mass effects in cross section calculations. These features will be explored in future work.           

The GMVFN scheme theory prediction for $Z+b$ at the LHC relies on PDF factorization~\cite{Collins:1989gx}, which holds at order $\alpha_s^3$ (NLO) and $\alpha_s^4$ (NNLO). It may be challenged at higher orders due to noncancellation of Glauber gluon exchanges involving the final-state HQ, which would require additional theory extensions.

Cross section measurements for $Z+c/b$ production in proton-proton collisions at the LHC will be delivered with high precision over a wide range of energies and for several kinematic distributions in the near future. They will play the central role in directly probing $c$ and $b$ PDFs over a wide kinematic domain. They can shed light on nonperturbative (intrinsic) HQ contributions in the nucleon as well as small-$x$ dynamics, and they are important for new physics searches. In fact, the $Z+b$ and $Z+b\bar b$ backgrounds are dominant in Higgs boson production in association with a $Z$ boson ($ZH$, $H\rightarrow b\bar b$) in the Standard Model, as well as in some processes beyond the Standard Model, such as production of SUSY Higgs bosons + $b$ quarks, or new generations of heavy quarks decaying into a $Z$ boson and a $b$ quark.

Future global PDF analyses are also going to be extended into a wide range of collision energies. These analyses are sensitive to HQ mass effects, including both phase-space suppression near the HQ mass threshold and large radiative corrections from collinear HQ production far above the threshold. Such effects are certainly comparable to NNLO and N$^3$LO corrections to hard-scattering cross sections. It is therefore essential that the fit evaluates all cross sections in a GMVFN scheme that varies the number of (nearly) massless quark flavors according to the typical energy and also includes heavy-quark mass dependence in the relevant kinematic regions. To achieve numerical accuracy, the same treatment of HQ mass effects must be adopted in the cross sections used to determine the PDFs and in new predictions based on these PDFs.

\subsection*{ACKNOWLEDGMENTS}
This work was performed in part at Aspen Center for Physics, which is supported by National Science Foundation Grant No. PHY-2210452. L.R. and D.W. thank the Galileo Galilei Institute for Theoretical Physics for the hospitality and the INFN for partial support during the completion of this work. The work of L.R. is supported in part by the U.S. Department of Energy under Grant No. DE- SC0010102.
M.G. is partially supported by the National Science Foundation under Grants No.~PHY-2112025 and No.~PHY-2412071.
P.M.N. was partially supported by the U.S. Department of Energy under Grant No.~DE-SC0010129 and thanks the U.S. DOE Institute for Nuclear Theory at the University of Washington for its hospitality during the completion of this work. The work of D.W. is supported in part by the National Science Foundation under Grants No.~PHY-2014021 and No.~PHY-2310363. K.X. was supported by the U.S. National Science Foundation under Grants No.~PHY-2310291 and PHY-2310497.

\subsection*{DATA AVAILABILITY}
No data were created or analyzed in this study.

%\clearpage 
%\newpage
\appendix

\bibliographystyle{utphys}
%\bibliography{ref}

\providecommand{\href}[2]{#2}\begingroup\raggedright\endgroup

\end{document}